\documentclass[
 aps,
 amsmath,amssymb,
 reprint,%
]{revtex4-1}
\usepackage{graphicx}% Include figure files
\usepackage{dcolumn}% Align table columns on decimal point
\usepackage{subfigure}
\usepackage{bm}% bold math
\usepackage{braket}
\usepackage[utf8]{inputenc}
\usepackage[T1]{fontenc}
\usepackage{mathptmx}
\usepackage{etoolbox}
\usepackage{caption}
\usepackage{textcomp, gensymb}

\usepackage{float}
\usepackage{xcolor}

\makeatletter
\def\@email#1#2{%
 \endgroup
 \patchcmd{\titleblock@produce}
  {\frontmatter@RRAPformat}
  {\frontmatter@RRAPformat{\produce@RRAP{*#1\href{mailto:#2}{#2}}}\frontmatter@RRAPformat}
  {}{}
}%
\makeatother

\usepackage{tikz,xcolor,hyperref}

\definecolor{lime}{HTML}{A6CE39}
\DeclareRobustCommand{\orcidicon}{
	\begin{tikzpicture}
	\draw[lime, fill=lime] (0,0) 
	circle [radius=0.16] 
	node[white] {{\fontfamily{qag}\selectfont \tiny ID}};
	\draw[white, fill=white] (-0.0625,0.095) 
	circle [radius=0.007];
	\end{tikzpicture}
	\hspace{-2mm}
}

\foreach \x in {A, ..., Z}{\expandafter\xdef\csname orcid\x\endcsname{\noexpand\href{https://orcid.org/\csname orcidauthor\x\endcsname}
			{\noexpand\orcidicon}}
}

\begin{document}

\preprint{AIP/123-QED}

\title[]{Alignment-based optically pumped magnetometer using a buffer gas cell}
% Force line breaks with \\

\author{L. M. Rushton\orcidE{}}
\homepage{Electronic mail: lucasrushton@outlook.com}
\author{L. Elson\orcidA{}}
\author{A. Meraki\orcidD{}}
\author{K. Jensen\orcidB{}}
\homepage{Electronic mail: kasper.jensen@nottingham.ac.uk}
\affiliation{School of Physics and Astronomy, University of Nottingham, University Park, Nottingham, NG7 2RD, UK}%

 %\homepage{http://www.Second.institution.edu/~Charlie.Author.}


%TC:ignore
\begin{abstract}
Alignment-based optically pumped magnetometers (OPMs) are capable of measuring oscillating magnetic fields with high sensitivity in the fT/$\sqrt{\text{Hz}}$ range. Until now, alignment-based magnetometers have only used paraffin-coated vapour cells to extend the spin relaxation lifetimes of the alkali vapour. The drawback of these cells is that they are hand-blown and are therefore time-intensive, and somewhat unreliable, to produce. Buffer gas cells, on the other hand, can be manufactured on a mass scale using microfabrication techniques. We present the first demonstration of an alignment-based magnetometer using a buffer gas vapour cell containing caesium (Cs) alkali vapour and nitrogen (N$_{\text{2}}$) buffer gas. The OPM is operated at $55\degree$C and we achieve a $325$~fT/$\sqrt{\text{Hz}}$ sensitivity to 10~kHz oscillating magnetic fields with an 800~Hz bandwidth. 
%We predict that we are close to the quantum-limited spin-projection noise. 
The alignment-based magnetometer uses a single laser beam for optical pumping and probing and could potentially allow for more rapid commercialisation of radio-frequency OPMs, due to the robustness of the one-beam geometry and the potential for mass-scale microfabrication of buffer gas cells.
\end{abstract}
%TC:endignore

%TC:ignore
\maketitle
%TC:endignore
\iffalse
\begin{quotation}
Lead paragraph
\end{quotation}
\fi

\section{Introduction}
Optically pumped magnetometers (OPMs) \cite{labyt_sander_tilmann_wakai_2022, auzinsh_budker_rochester_2014, budker_romalis_2007} based on spin-polarized atoms (e.g.~alkali atoms such as caesium (Cs) or rubidium) can measure magnetic fields with high sensitivity in the fT/$\sqrt{\text{Hz}}$ range \cite{kominis_kornack_allred_romalis_2003, Wasilewski2010, Chalupczak2012, yao_maddox_renzoni_2022}. 
Current commercial OPMs \cite{quspin, fieldline, twinleaf} are operated close to zero magnetic field %($B_{0}=0$) 
in the spin-exchange relaxation-free (SERF) regime measuring one, two or three components of the magnetic field, or in the Earth's field as scalar magnetometers measuring the total magnetic field amplitude. 
These OPMs use one or two beams of circularly polarized light generated from a single laser diode inside the OPM, making the sensors compact and robust. The circularly polarized light effectively generates spin-orientation along the light propagation (i.e.,~the atomic spins point in a certain direction) which responds to magnetic fields and can be measured by detecting the transmitted light.
When detecting oscillating magnetic fields in the kHz-MHz frequency range, radio-frequency (RF) OPMs \cite{savukov_seltzer_romalis_sauer_2005, auzinsh_budker_rochester_2014, rochester_thesis, deans_marmugi_renzoni_2018_subpT_unshielded_tabletop, deans_cohen_yao_maddox_vigilante_renzoni_2021, dhombridge_claussen_iivanainen_schwindt_2022} must be used. One type of RF OPM using only a single laser beam is the alignment-based magnetometer \cite{ledbetter_acosta_rochester_budker_pustelny_yashchuk_2007, zigdon_wilson-gordon_guttikonda_bahr_neitzke_rochester_budker_2010, atomicdensitymatrix_nonlinear}, which uses linearly polarized light capable of effectively aligning the atoms in the direction perpendicular to its propagation. As a result, as the RF field affects such alignment being created, its presence can be sensed directly by measuring properties of the same beam.

High sensitivity optical magnetometry requires a long atomic spin-coherence time. This can be achieved using vapour cells coated on the inside with an anti-relaxation coating (e.g.~paraffin), such that the moving alkali atoms can bounce off the inner glass walls of the vapour cell many times  without losing their spin-coherence \cite{Balabas2010prl, Li2017}.
Alternatively, a long coherence time can be achieved by filling the vapour cell with buffer gas (e.g.~N$_{\text{2}}$). Rapid collisions between the buffer gas atoms and the alkali atoms make the alkali atoms diffuse slowly, which mitigates the effects of spin-destroying wall collisions.
Alkali vapour cells for magnetometry are typically hand-blown, however buffer gas cells for magnetometry can be produced on a mass scale using microfabrication techniques \cite{Shah2007natphot, Dyer_2022}. Such microfabrication techniques have not, as of yet, been compatible with anti-relaxation coating.

So far, alignment-based optical magnetometry has been demonstrated using hand-blown, anti-relaxation coated cells \cite{ledbetter_acosta_rochester_budker_pustelny_yashchuk_2007, zigdon_wilson-gordon_guttikonda_bahr_neitzke_rochester_budker_2010}.
The presence of buffer gas leads to pressure broadening of the alkali vapour absorption spectrum, reducing the light-atom coupling and affecting the optical pumping preparing the aligned state. The buffer gas N$_{\text{2}}$ is also a quenching gas \cite{seltzer_thesis} which causes the alkali atoms not to de-excite via spontaneous emission. Rapid collisional mixing in the excited state \cite{seltzer_thesis} also occurs in buffer gas cells, but not in paraffin-coated cells. We show here that, despite these complexities, it is possible to realise an alignment-based magnetometer using a buffer gas cell. 
We experimentally demonstrate an alignment-based magnetometer using a Cs alkali vapour and 65~Torr N$_{\text{2}}$ buffer gas cell with a sensitivity of $325$~fT/$\sqrt{\text{Hz}}$ to oscillating magnetic fields at 10~kHz. We also demonstrate an alignment-based magnetometer with a paraffin-coated cell placed in the same experimental setup to verify the methods and for comparison. Our results open up the possibility for miniaturisation \cite{dhombridge_claussen_iivanainen_schwindt_2022, rushton_2022} and commercialisation of RF OPMs, with potential impact in areas such as medical physics \cite{deans_marmugi_renzoni_2020, jensen_zugenmaier_arnbak_staerkind_balabas_polzik_2019, marmugi_renzoni_2016}, remote sensing \cite{rushton_2022, deans_marmugi_renzoni_2018} and non-destructive testing \cite{bevington_gartman_chalupczak_2021, bevington_gartman_chalupczak_2019}.

\section{Alignment-based optical magnetometry}
\begin{figure}
\includegraphics[width=\linewidth]{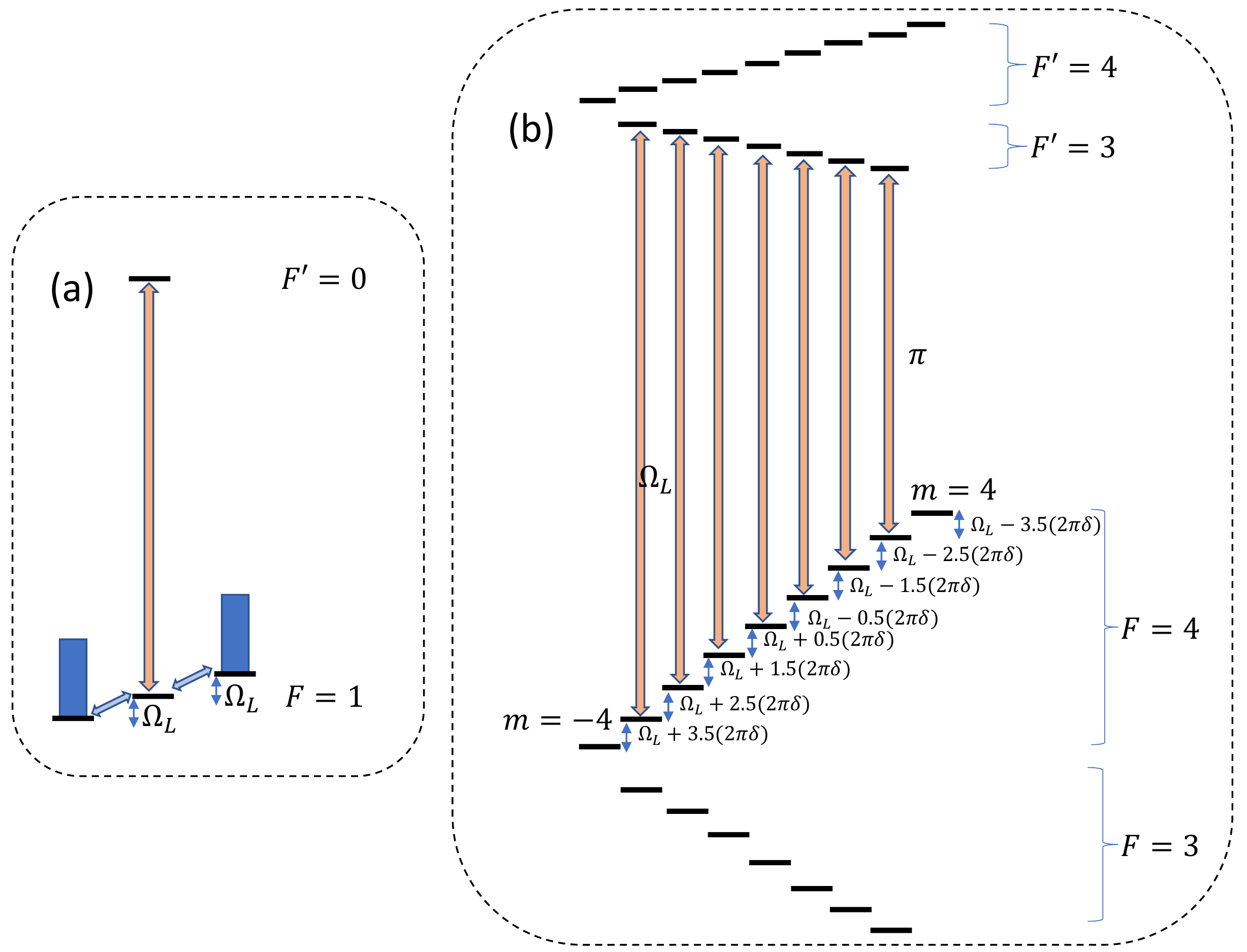}
\caption{ (a) Energy level diagram with a $F=1$ ground state and $F'=0$ excited state. 
The neighbouring groundstate sublevels are each split by the Larmor frequency $\Omega_{L}$.
%At ``small'' $\textbf{B}_{0}$ the neighbouring sublevels are each split by $\Delta E=\hbar\Omega_{L}$. 
Linearly polarised ($\pi$) light drives transitions between $m=0$ and $m'=0$. (b) D1 line of Cs with $F=3,4$ ground states and $F'=3,4$ excited states. Linearly polarised ($\pi$) light drives transitions between $F=4$ and $F'=3$. Small corrections ($\delta$) for the non-linear Zeeman splitting are shown. \label{fig:Threelevels}}
\end{figure}

The theory underpinning the alignment-based magnetometer
%is explained in detail in the Refs.~
\cite{auzinsh_budker_rochester_2014, zigdon_wilson-gordon_guttikonda_bahr_neitzke_rochester_budker_2010, rochester_thesis, atomicdensitymatrix_nonlinear} 
will now be revised and discussed.
Consider atoms with a  $F=1\rightarrow F'=0$ optical transition
with groundstate sublevels
$|F,m \rangle =
\left\{  |1,1 \rangle, |1,0 \rangle, |1,-1 \rangle  \right\}$ 
and an excited state $|F',m' \rangle= |0',0' \rangle $  
as shown in Fig.~\ref{fig:Threelevels}(a). A single laser beam is used for optical pumping and probing of the atoms. Assume the light is linearly ($\pi$) polarized along the direction of a static magnetic field $B_0\hat{\textbf{z}}$.
In this case, the atoms will be optically pumped into the $m=\pm1$ sublevels with equal probability, creating a so-called ``spin-aligned state''. This is a dark state, such that with perfect optical pumping, the light will be fully transmitted through the atomic vapour.
Now assume further that there is a transverse oscillating (RF) magnetic field which we would like to detect. That RF field will affect the optical pumping and thereby the transmitted light which can be detected by measuring its intensity or polarization.

The total Hamiltonian $\hat{H}$ which describes the system is given by 
\begin{equation}
    \hat{H} = \hat{H}_{0} + \hat{H}_{l} + \hat{H}_{B},
\end{equation}
where $\hat{H}_{0}$, $\hat{H}_{l}$ and $\hat{H}_{B}$ are the unperturbed, light-atom interaction and magnetic field-atom interaction Hamiltonians, respectively. 
The unperturbed Hamiltonian $\hat{H}_{0}$ written in the overall basis 
$\{\ket{1,1} \equiv \ket{1}, \ket{1,0} \equiv \ket{0}, \ket{1,-1} \equiv \ket{-1}, \ket{0',0'} \equiv \ket{0'}\}$ is
\begin{equation}
\begin{split}
\hat{H}_{0}=
%&~
\begin{pmatrix}
0 & 0 & 0 & 0\\
0 & 0 & 0 & 0\\
0 & 0 & 0 & 0\\
0 & 0 & 0 & \hbar \omega_{0}
\end{pmatrix}
\end{split}
\end{equation}
where $\hbar$ is the reduced Planck constant, $\omega_{0}=2\pi c /\lambda$ is the optical transition frequency, $\lambda$ its wavelength and $c$ the speed of light. 
The light-atom interaction is governed by 
\begin{equation}
    \hat{H}_{l} = -\textbf{E}\cdot\hat{\textbf{d}},
\end{equation}
where $\hat{\textbf{d}}$ is the dipole operator and $\textbf{E} = E_{0} \cos (\omega t) \hat{\textbf{z}}$ is the electric field of the light. The light-atom interaction Hamiltonian $\hat{H}_{l}$ is
\begin{equation}
    \hat{H}_{l} = \hbar \Omega_{R} \cos \omega t \begin{pmatrix}
    0 & 0 & 0 & 
    0 \\ 
    0 & 0 & 0 & 
    -1 \\ 
    0 & 0 & 0 & 
    0 \\ 
    0 & -1 & 0 & 
    0
    \end{pmatrix},
\end{equation}
where $\Omega_{R} = \braket{1||d||0'}E_{0}/(\sqrt{3}\hbar)$ is the Rabi frequency and $\braket{1||d||0'}$ is the transition dipole matrix element. Assuming $B_{x}=B_{\text{RF}}\cos\omega_{\text{RF}}t$, $B_{y}=0$ and $B_{z}=B_{0}$, the magnetic field-atom interaction is given by 
\begin{equation}
\begin{split}
    \hat{H}_{B} & = -\hat{\bm{\mu}}\cdot\textbf{B} = \frac{g_{F} \mu_{B}}{\hbar}(\hat{F}_{x}B_{x} + \hat{F}_{z}B_{z}) 
\\ & = g_{F}\mu_{B}\begin{pmatrix}
    B_{0} & \frac{B_{\text{RF}}\cos \omega_{\text{RF}} t}{\sqrt{2}} & 0 & 0\\
    \frac{B_{\text{RF}}\cos \omega_{\text{RF}} t}{\sqrt{2}} & 0 & \frac{B_{\text{RF}}\cos \omega_{\text{RF}} t}{\sqrt{2}} & 0\\
    0 & \frac{B_{\text{RF}}\cos \omega_{\text{RF}} t}{\sqrt{2}} & -B_{0} & 0\\
    0 & 0 & 0 & 0
    \end{pmatrix}.
    \end{split}
\end{equation}
Here  $\hat{\bm{\mu}}=g_{F}\mu_{B}(\hat{F}_{x}\hat{\textbf{x}}+\hat{F}_{y}\hat{\textbf{y}}+\hat{F}_{z}\hat{\textbf{z}})/\hbar$ is the Cs atom's magnetic dipole operator, $g_F$ is the hyperfine Lande g-factor \cite{steck}, and $\mu_B$ is the Bohr magneton.
Defining the Larmor frequency $ \Omega_{L}=g_{F}\mu_{B}B_{0}/\hbar$ and letting the strength of the RF field be represented by $\Omega_{\text{RF}}=g_{F}\mu_{B} B_{\text{RF}}/\hbar$, the total Hamiltonian $\hat{H}=\hat{H}_{0}+\hat{H}_{B}+\hat{H}_{l}$ is

\begin{equation}
\begin{split}
    \hat{H} = \hbar\begin{pmatrix}
    \Omega_{L} & \frac{\Omega_{\text{RF}}\cos \omega_{\text{RF}} t}{\sqrt{2}} & 0 & 0\\
    \frac{\Omega_{\text{RF}}\cos \omega_{\text{RF}} t}{\sqrt{2}} & 0 & \frac{\Omega_{\text{RF}}\cos \omega_{\text{RF}} t}{\sqrt{2}} & -\frac{\Omega_{R}\cos\omega t}{\sqrt{3}}\\
    0 & \frac{\Omega_{\text{RF}}\cos \omega_{\text{RF}} t}{\sqrt{2}} & -\Omega_{L} & 0\\
    0 & -\frac{\Omega_{R}\cos\omega t}{\sqrt{3}} & 0 & \omega_{0}
    \end{pmatrix}.
    \end{split}
    \label{eq:Hamiltonian_LabFrame}
\end{equation}
Going to a rotating frame at the optical frequency $\omega$, followed by going to another rotating frame at the RF frequency $\omega_{\text{RF}}$, then neglecting the fast oscillating terms using the rotating wave approximation and setting $\Delta = \omega - \omega_{0}$, $\Delta_{\text{RF}}=\omega_{\text{RF}}-\Omega_{L}$, the Hamiltonian $\hat{\Tilde{H}}$ in the rotating frame is 
\begin{equation}
    \begin{split}
        \hat{\Tilde{H}} = \hbar\begin{pmatrix}
    -\Delta_{\text{RF}} & \frac{\Omega_{\text{RF}}}{2\sqrt{2}} & 0 & 0\\
    \frac{\Omega_{\text{RF}}}{2\sqrt{2}} & 0 & \frac{\Omega_{\text{RF}}}{2\sqrt{2}} & -\frac{\Omega_{R}}{2\sqrt{3}}\\
    0 & \frac{\Omega_{\text{RF}}}{2\sqrt{2}} & \Delta_{\text{RF}} & 0\\
    0 & -\frac{\Omega_{R}}{2\sqrt{3}} & 0 & -\Delta
    \end{pmatrix}.
    \end{split}
\end{equation}
Next the relaxation $\hat{\Gamma}$ and repopulation $\hat{\Lambda}$ matrices must be taken into account and are given by
\begin{equation}
    \hat{\Gamma} = \begin{pmatrix}
    \gamma & 0 & 0 & 0\\
    0 & \gamma & 0 & 0\\
    0 & 0 & \gamma & 0\\
    0 & 0 & 0 & \gamma + \Gamma
    \end{pmatrix},
\end{equation}

\begin{equation}
        \hat{\Lambda} = \begin{pmatrix}
        \frac{\gamma + \Gamma \rho_{0',0'}}{3} & 0 & 0 & 0\\
        0 & \frac{\gamma + \Gamma \rho_{0',0'}}{3} & 0 & 0\\
        0 & 0 & \frac{\gamma + \Gamma \rho_{0',0'}}{3} & 0\\
        0 & 0 & 0 & 0
        \end{pmatrix}.
\end{equation}
The atoms in the excited state decay to the ground state sublevels at a rate $\Gamma$ and are assumed to repopulate the three ground state sublevels in equal proportion $\Gamma \rho_{0', 0'}/3$, where $\rho_{0', 0'}$ is the population of the excited state.
For a paraffin-coated cell, the excited atoms decay via spontaneous emission (at the rate $2\pi (4.56~\text{MHz})$ \cite{steck} for Cs atoms in the first excited state).
For a buffer gas cell, the excited atoms mainly decay via quenching, as discussed in the next section.
Note that the model above only includes one excited state and therefore does not describe collisional mixing (between multiple excited states).
The atoms also have a spin-coherence (or transverse relaxation) time $T_{2}=1/\gamma$.
In a buffer gas cell, the alkali atoms diffuse slowly due to collisions with the buffer gas, increasing $T_{2}$. The spins relax when the alkali atoms hit the glass walls due to electron randomisation collisions or via spin-exchange or spin-destruction collisions  between two alkali atoms \cite{graf_kimball_rochester_kerner_wong_budker_alexandrov_balabas_yashchuk_2005, ledbetter_acosta_rochester_budker_pustelny_yashchuk_2007, labyt_sander_tilmann_wakai_2022}.
In a paraffin-coated cell the alkali atoms can bounce off the walls thousands of times before spin relaxation occurs \cite{Balabas2010prl}.
Power broadening due to laser light also reduces $T_2$.

The Liouville equation for the density matrix $\hat{\Tilde{\rho}}$ in the rotating frame is given by 
\begin{equation}
i \hbar \frac{\partial \hat{\Tilde{\rho}}}{\partial t} = [\hat{\Tilde{H}}, \hat{\Tilde{\rho}}] - i\hbar \frac{1}{2}(\hat{\Gamma}\hat{\Tilde{\rho}} + \hat{\Tilde{\rho}}\hat{\Gamma}) + i \hbar \hat{\Lambda}.
\end{equation}
In the steady state $d\hat{\Tilde{\rho}}/dt=0$ and the right-hand-side of the equation can be solved to determine $\hat{\Tilde{\rho}}$ in the rotating frame. The density matrix is returned to the lab frame $\hat{\rho}$ by using a transformation matrix. The polarisation $\textbf{P}=n\text{Tr}(\hat{\rho}\hat{\textbf{d}})$ of the atomic vapour can then be calculated, where $n$ is the alkali atom number density. The formalism in Ref.~\cite{auzinsh_budker_rochester_2014} allows for the in-phase and out-of-phase rotations of a linearly polarised beam to be extracted \cite{zigdon_wilson-gordon_guttikonda_bahr_neitzke_rochester_budker_2010, atomicdensitymatrix_nonlinear}. The expressions for the in-phase $ \partial\phi^{\text{in}}/ \partial l$ and quadrature $ \partial\phi^{\text{out}}/ \partial l$ values are \cite{zigdon_wilson-gordon_guttikonda_bahr_neitzke_rochester_budker_2010}
%%%%%%%%%%%%%%%%%%%%%%%%%%%%%%%%%%%%%%
\begin{align}
    \frac{\partial\phi^{\text{in}}}{\partial l} & = \frac{n \Delta_{\text{RF}}\lambda^{2}\Omega_{\text{RF}}(2\gamma^{2}+8\Delta_{\text{RF}}^{2}-\Omega_{\text{RF}}^{2})\Omega_{R}^{2}}{36\pi\Gamma\gamma(\gamma^{2}+4\Delta_{\text{RF}}^{2}+\Omega_{\text{RF}}^{2})[4(\gamma^{2}+\Delta_{\text{RF}}^{2})+\Omega_{\text{RF}}^{2}]} \nonumber \\ 
     & \approx \frac{n \lambda^2}{72 \pi} \cdot 
     \frac{\Omega_R^2}{ \Gamma}     \cdot
     \Omega_{\text{RF}}    \cdot
     \frac{\Delta_{\text{RF}}/\gamma}{\Delta_{\text{RF}}^{2} + \gamma^2} 
    \quad \mathrm{for} \; \; \Omega_{\text{RF}}^{2}\ll \gamma^{2},
   \label{eq:Zigdon_inphase_new}
\end{align}
\begin{align}
    \frac{\partial\phi^{\text{out}}}{\partial l} & = \frac{n \lambda^{2}\Omega_{\text{RF}}(4\gamma^{2}+16\Delta_{\text{RF}}^{2}+\Omega_{\text{RF}}^{2})\Omega_{R}^{2}}{72\pi\Gamma(\gamma^{2}+4\Delta_{\text{RF}}^{2}+\Omega_{\text{RF}}^{2})[4(\gamma^{2}+\Delta_{\text{RF}}^{2})+\Omega_{\text{RF}}^{2}]} \nonumber \\
        & \approx \frac{n \lambda^2}{72 \pi} \cdot 
     \frac{\Omega_R^2}{\Gamma}     \cdot
     \Omega_{\text{RF}}    \cdot
     \frac{1}{\Delta_{\text{RF}}^{2} + \gamma^2} 
     \quad 
     \mathrm{for} \; \; \Omega_{\text{RF}}^{2}\ll \gamma^{2} , 
   \label{eq:Zigdon_outofphase_new}
\end{align}
where $l$ is the length of the vapour cell. In the limit when $\Omega_{\text{RF}}^{2}\ll \gamma^{2}$, as is the case throughout this paper, 
$\partial\phi^{\text{in}}/\partial l$ and  $\partial\phi^{\text{out}}/\partial l$ 
are proportional to the RF magnetic field amplitude $B_{\text{RF}} \propto \Omega_{\text{RF}}$ and 
have  dispersive- and absorptive-Lorentzian lineshapes, respectively, when varying the RF detuning $\Delta_{\text{RF}}$.
The light polarization rotation is measured using a balanced photodetector and lock-in detection (at the RF frequency), yielding the lock-in outputs which can be written as
\begin{align}
 X  \propto &
\frac{\partial\phi^{\text{out}}}{\partial l} 
\propto 
B_{\text{RF}} \cdot \frac{1}{\left( \omega_{\text{RF}}-\Omega_{L} \right)^{2} + \gamma^2} , \label{eq:X} \\
Y \propto & 
\frac{\partial\phi^{\text{in}}}{\partial l}
\propto 
B_{\text{RF}} \cdot \frac{\left( \omega_{\text{RF}}-\Omega_{L} \right)/\gamma}{\left( \omega_{\text{RF}}-\Omega_{L} \right)^{2} + \gamma^2} ,  \label{eq:Y} \\
R = & 
\sqrt{X^2+Y^2} = |X+iY| \propto 
B_{\text{RF}} \cdot \left|\frac{1+i\left( \omega_{\text{RF}}-\Omega_{L} \right)/\gamma}{\left( \omega_{\text{RF}}-\Omega_{L} \right)^{2} + \gamma^2} \right|.
\label{eq:R}
\end{align}

\section{Caesium}
\subsection{Optical pumping in a paraffin-coated cell}
\begin{figure*}
\subfigure[]{\label{fig:}\includegraphics[width=0.49\textwidth]{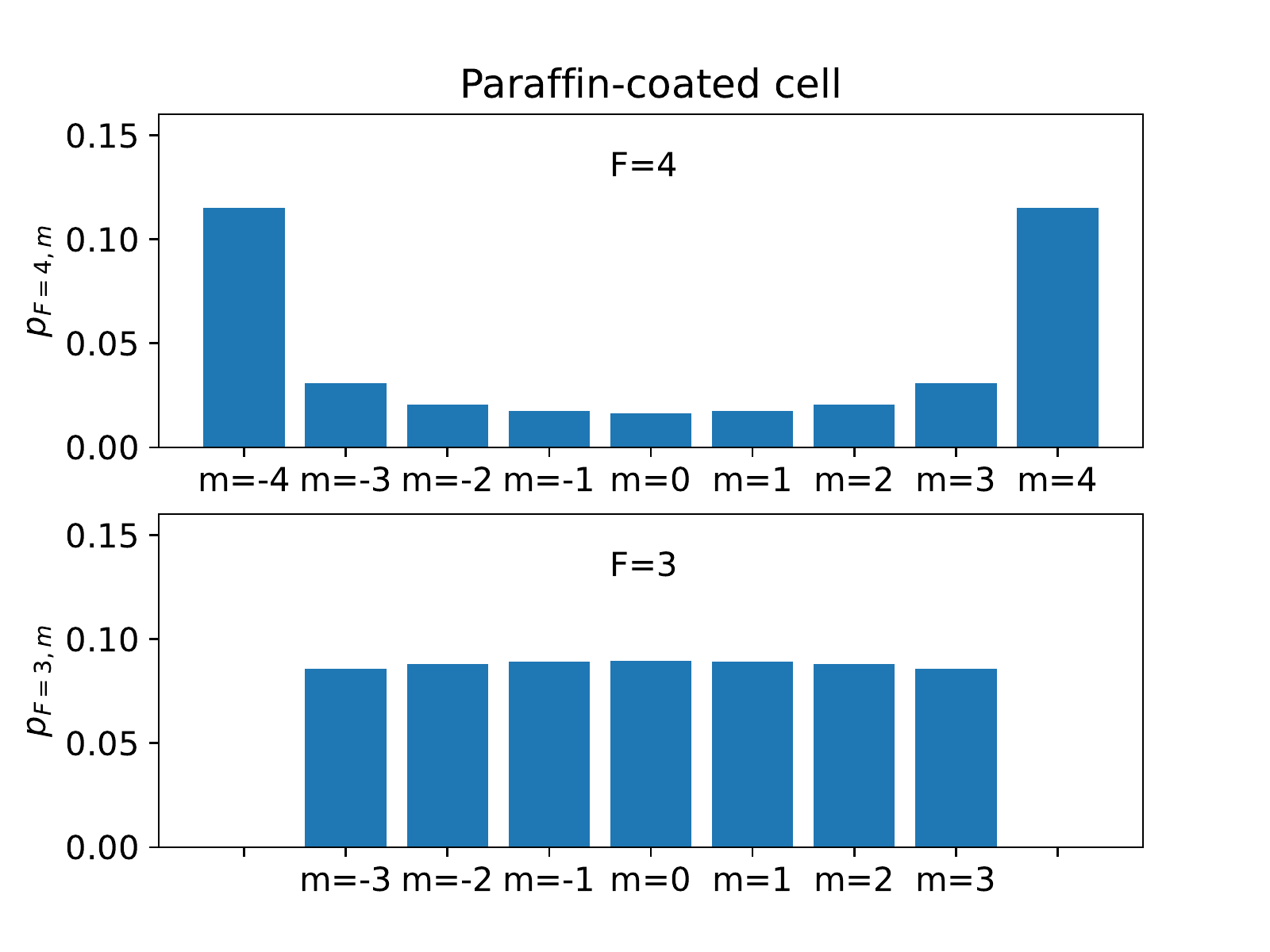}}
\subfigure[]{\label{fig:}\includegraphics[width=0.49\textwidth]{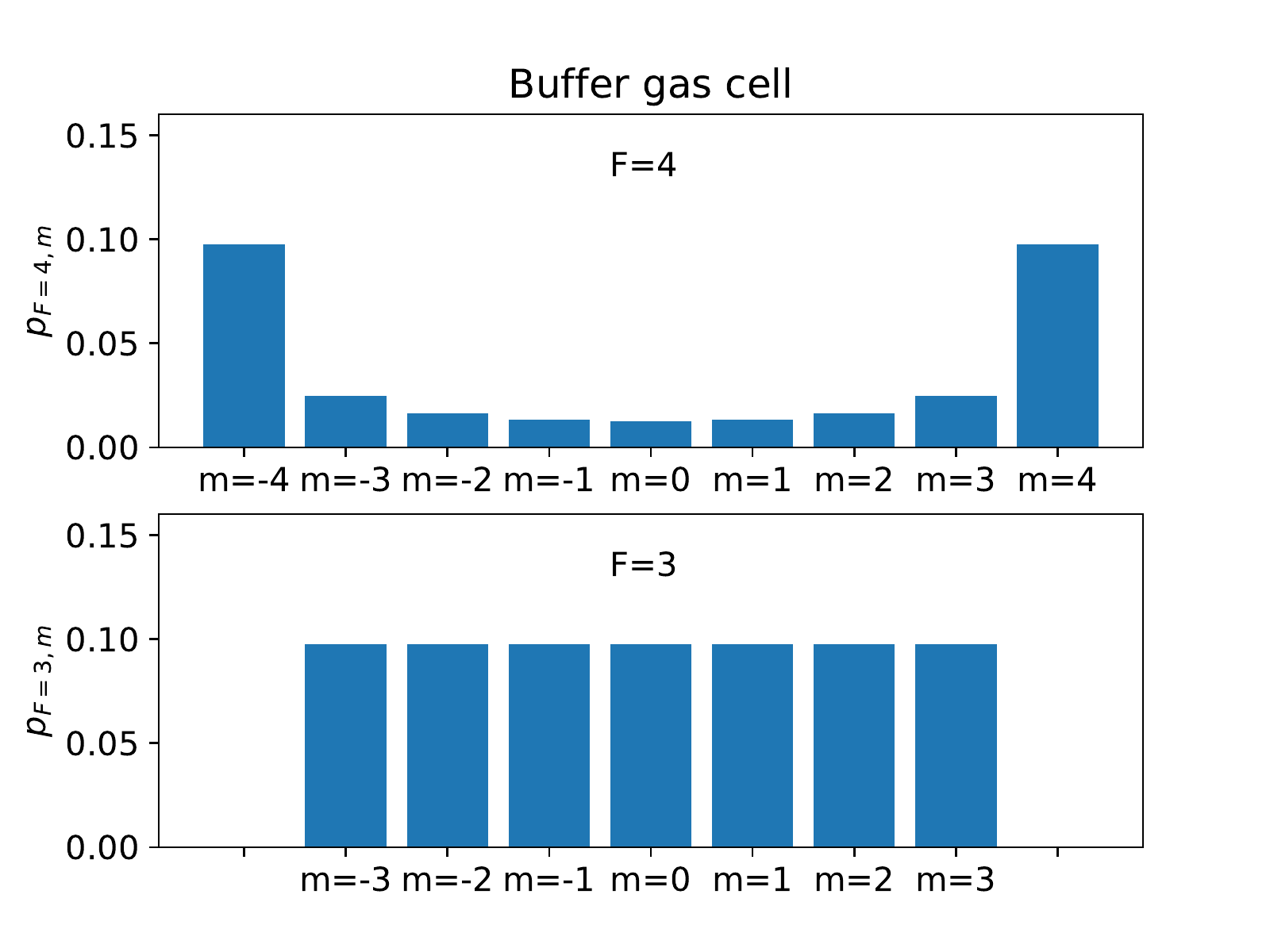}}
\caption{\label{fig:chapter02_Pi_OPvstime_BarGraph_AM} Optical pumping from $F=4\rightarrow F'=3$ with $\pi$-polarised light. The populations of the $F=3$ and $F=4$ ground state magnetic sublevels in the steady state are plotted, with a longitudinal relaxation rate $\Gamma_{1}=R_{p}/20$ for (a) a paraffin-coated cell where the dominant de-excitation mechanism from the excited state is spontaneous emission, and (b) a buffer gas cell where the dominant de-excitation mechanism is quenching.}
\end{figure*}

A caesium (Cs) atom \cite{steck} has two ground states with hyperfine quantum numbers $F=3$ and $F=4$ separated by the hyperfine splitting 
$ \nu_{\text{hf}}=9.192$~GHz. 
The first excited states have $F'=3$ and $F'=4$ with a hyperfine splitting of 1.2~GHz.  
The optical transition of interest for this experiment is the Cs D1 $F=4\rightarrow F'=3$ transition using $z$-linearly polarised light (see Fig.~\ref{fig:Threelevels}b) with a wavelength around 895~nm. 
The optical pumping can be understood by determining the populations of the 16 magnetic sublevels of the Cs ground state using rate equations \cite{eckel2022}. We will first consider a paraffin-coated cell, where the decay from the excited state is by spontaneous emission. An example of the rate of change of the population of the magnetic sublevel $F=3, m=3$, $dp_{\text{F=3,m=3}}/dt$, i.e., the diagonal element of the density matrix, is

\begin{equation}
\begin{split}
    \frac{dp_{\text{4,3}}}{dt} =&~R_{p}(-p_{\text{4,3}}c_{4,3\leftrightarrow3',3'} + p_{\text{4,2}}  c_{4,2\leftrightarrow3',2'}  c_{4,3\leftrightarrow3',2'} \\ & +
                    p_{\text{4,3}}  c_{4,3\leftrightarrow3',3'}  c_{4,3\leftrightarrow3',3'}) - \Gamma_{1}  p_{4,3} + \Gamma_{1}/16,
\end{split}
\label{eq:OpticalPumping_Pi_EgDifferential}
\end{equation}
where $R_{p}$ is the optical pumping rate, $p_{4, 3}=p_{4,3}(t)$ is the population of the magnetic sublevel at time $t$, and $c_{4,2\leftrightarrow 3',2'}$ is the Clebsch-Gordon coefficient squared \cite{steck} for the $\pi$ transition from $F=4,m=2 \leftrightarrow F'=3, m'=2$, for example. The longitudinal relaxation rate $\Gamma_{1}$ was not measured experimentally in this work but is typically much smaller than the transverse relaxation rate $\gamma$ \cite{julsgaard_2003}. The negative terms in Eq.~\ref{eq:OpticalPumping_Pi_EgDifferential} depopulate the magnetic sublevel and the positive terms repopulate the sublevel. The populations of the 16 magnetic sublevels in the $F=3$ and $F=4$ ground states in the steady state ($dp/dt=0$) are determined numerically and we plot these in Fig.~\ref{fig:chapter02_Pi_OPvstime_BarGraph_AM}a. 
We see that the atoms in the $F=4$ sublevels are symmetrically distributed with most atoms in the $m=\pm4$ sublevels, corresponding to a spin-aligned state.
Note that many atoms are ``lost'' to the other ground state $F=3$. These atoms are not probed as they are 9.192~GHz detuned from the light. 

\subsection{Optical pumping in a buffer gas cell}
If a buffer gas such as 65~Torr of N$_{\text{2}}$ is present in a Cs vapour cell without any paraffin coating, then the Cs atoms will mostly decay via quenching rather than via spontaneous emission \cite{seltzer_romalis_2009, seltzer_thesis}, as will now be shown. The many vibrational and rotational states of the quenching gas molecule, in this case N$_{\text{2}}$, mean that when a Cs atom in the excited state collides with a N$_{\text{2}}$ molecule, the Cs atom can de-excite without the emission of a photon, instead transferring its energy to the many vibrational and rotational modes of the N$_{\text{2}}$ molecule. The quenching rate $R_{Q}$ is given by
\begin{equation}
    R_{Q}=n_{Q}\sigma_{Q}v_{\text{Cs,N}_{2}},
\end{equation}
where $n_{Q}=P/(k_{B}T)=1.91\times10^{24}$~m$^{-3}$ is the number density of N$_{\text{2}}$ molecules at $T\sim55\degree$C, 
$P$ is the pressure, $k_B$ is the Boltzmann constant,
$\sigma_{Q}=5.5\times10^{-19}$~m$^{2}$ \cite{seltzer_thesis} is the quenching gas cross-section for Cs and N$_{\text{2}}$ (at 100$\degree$C) and $v_{\text{Cs,N}_{2}}=\sqrt{8k_{B}T/\pi M}=548$~m/s is the relative velocity between a Cs atom and N$_{\text{2}}$ molecule. The mass $M=3.84\times10^{-26}$~kg is the effective mass of a Cs atom and N$_{\text{2}}$ molecule, given by 
$M=m_{\text{Cs}}m_{\text{N}_{2}}/\left(m_{\text{Cs}}+m_{\text{N}_{2}}\right)$.
The quenching factor $Q$ helps determine the dominant decay mechanism, whether by spontaneous emission ($Q=1$) or by quenching ($Q=0$), and is given by \cite{seltzer_thesis}
\begin{equation}
Q = \frac{1}{1+R_{Q}\tau_{\text{nat}}}.
\end{equation}
Calculating $R_{Q}=5.9\times10^{8}$~s$^{-1}$ from the parameters stated above for Cs and 65~Torr N$_{\text{2}}$ and taking the natural lifetime of the D1 excited state to be $\tau_{\text{nat}}=35$~ns, then $Q=0.05$. This means that, for the 65~Torr N$_{\text{2}}$ buffer gas cell used in our experiments, the dominant de-excitation mechanism from the excited state is quenching. During quenching, the decay probabilities to the ground states are not governed by the Clebsch-Gordon coefficients. Instead the atoms decay with equal probability (1/16) to any of the $F=3$ and $F=4$ ground state magnetic sublevels. Crucially, though, the $F=4, m=\pm4$ states will still be dark states in the presence of N$_{\text{2}}$, a quenching gas. An example of a rate equation for the population $dp_{4,3}/dt$ of the $F=4, m=3$ magnetic sublevel is given by 
\begin{equation}
\begin{split}
    \frac{dp_{4,3}}{dt}=&~R_p (- p_{4,3}  c_{4,3\leftrightarrow3',3'} + \frac{1}{16}[p_{4,3} c_{4,3\leftrightarrow3',3'} \\&~+ p_{4,2}  c_{4,2\leftrightarrow3',2'} 
                 + p_{4,1}  c_{4,1\leftrightarrow3',1'} + p_{4,0}  c_{4,0\leftrightarrow3',0'} 
                 \\&~+ p_{4,-1}  c_{4,-1\leftrightarrow3',-1'} + p_{4,-2}  c_{4,-2\leftrightarrow3',-2'}
                 \\&~+ p_{4,-3}  c_{4,-3\leftrightarrow3',-3'}]) - \Gamma_{1}  p_{4,3} + \frac{\Gamma_{1}}{16}.
\end{split}
\label{eq:rateBuffer}
\end{equation}
The 16 rate equations are solved in the steady state and an illustrative example of optical pumping with a buffer gas is shown in Fig.~\ref{fig:chapter02_Pi_OPvstime_BarGraph_AM}b. 
We see that the distribution of atoms in the ground state sublevels is similar for both buffer gas and paraffin-coated vapour cells.
It is assumed that $Q=0$, which is a safe assumption to make for the 65~Torr N$_{\text{2}}$ buffer gas cell in this paper. 
In the above we assumed that the excited states $F'=3$ and $F'=4$ are resolved such that the light is only resonant with the $F=4 \rightarrow F'=3$ transition.
We note, however, that the $F'=4$ excited state would need to be incorporated into the rate equations if the buffer gas pressure becomes significantly larger, as discussed in Sec.~\ref{sec:buffer}.

\subsection{Non-linear Zeeman splitting}
A Cs atom in the $F=4$ ground state has $2F+1=9$ sublevels $|F,m\rangle$ which, when placed in a small magnetic field $B_0$, have the energy $E(m)=m h \nu_L$ due to the linear Zeeman effect. Here $\nu_L$ is the Larmor frequency in Hz. That is to say, the splittings between neighbouring sublevels are all equal to the Larmor frequency 
$\Delta \nu_{m,m-1} \equiv 
\left( E(m)-E(m-1) \right)/h
= \nu_L$. 
In this case, a single magnetic resonance will be observed when sweeping the RF frequency $\nu_{\text{RF}}$ (in Hz) across the Larmor frequency $\nu_L$ and measuring the polarization rotation of the transmitted light (see 
Eq.~\ref{eq:X}, \ref{eq:Y} and \ref{eq:R}).
However, at larger magnetic fields, the splittings between sublevels are slightly different due to the non-linear Zeeman effect.
We calculate \cite{julsgaard_2003, bao_wickenbrock_rochester_zhang_budker_2018, steck}
\begin{equation}
\Delta \nu_{m,m-1} = \nu_L - \delta \left( m-\frac{1}{2}\right),
\end{equation}
where the non-linear Zeeman splitting (in Hz) is
\begin{equation}
\delta = \frac{2\nu_L^2}{\nu_{\text{hf}}}
\label{eq:NZS0}
\end{equation}
as illustrated in Fig.~\ref{fig:Threelevels}(b). 
In particular, the difference in transition frequencies between $\Delta \nu_{4,3}$ and $\Delta \nu_{-3, -4}$ is
\begin{equation}
\left| \Delta \nu_{4,3} - \Delta \nu_{-3, -4}  \right| = 7 \delta.
\label{eq:NZS}
\end{equation}
In other words, at larger magnetic fields a total of 8 magnetic resonances should be observed when sweeping the RF field across the Larmor frequency with the outermost resonances split by $7\delta$.

\section{Paraffin-coated cell}
\begin{figure}[ht]
\subfigure[]{\label{fig:alignment_magnetometer_setup}\includegraphics[width=0.49\textwidth]{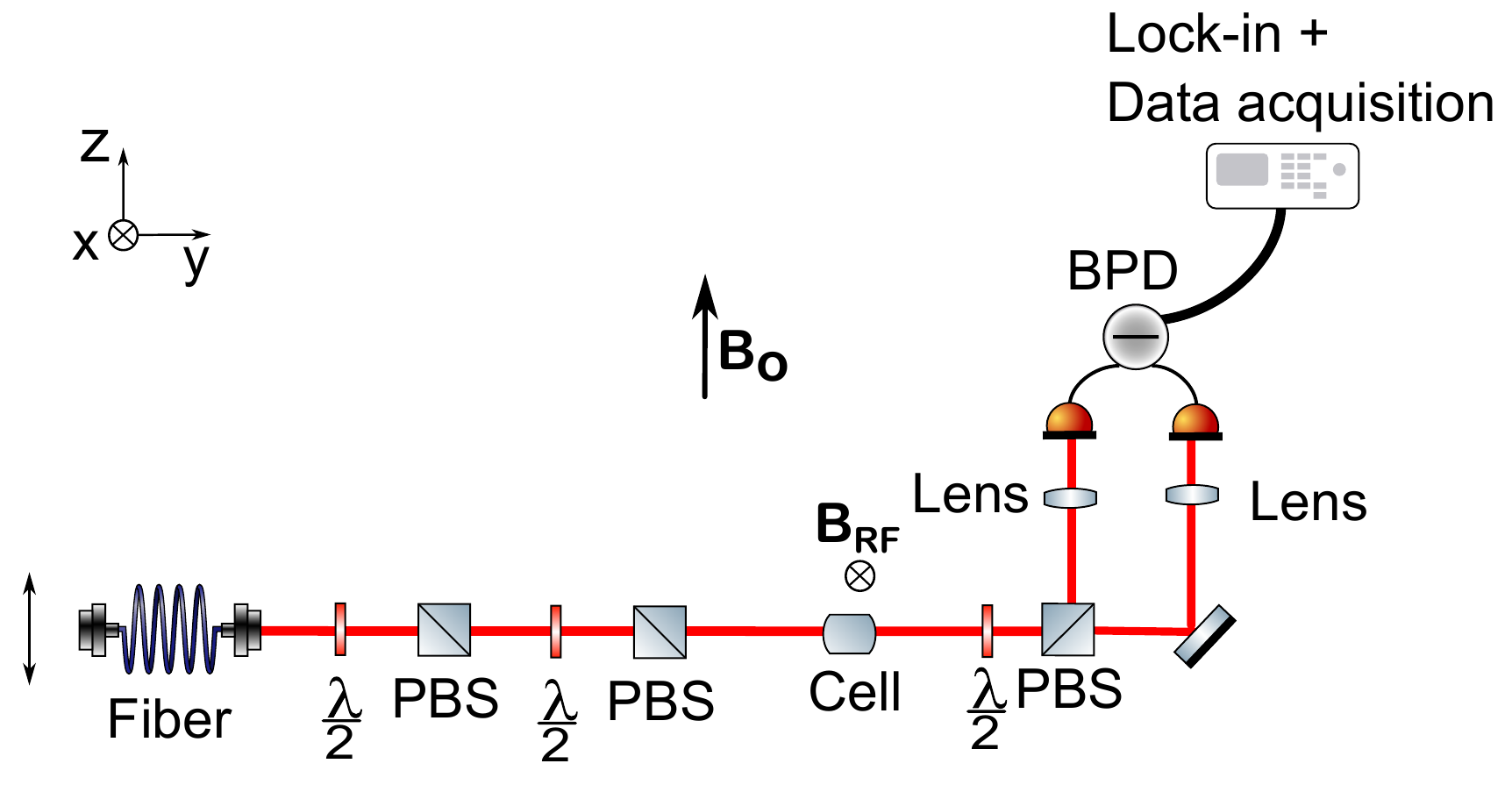}}
\subfigure[]{\label{fig:cellA}\includegraphics[width=0.155\textwidth]{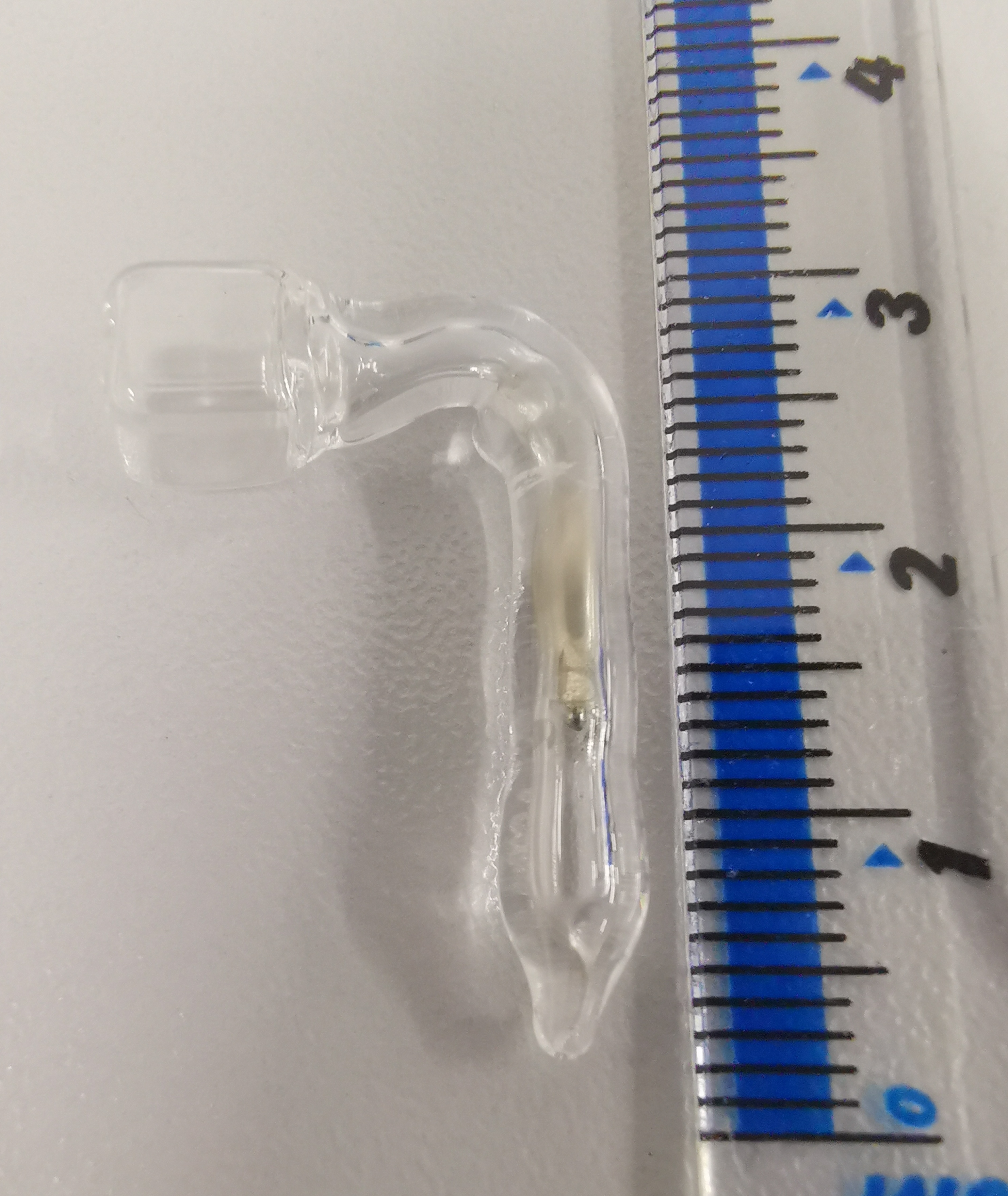}}
\subfigure[]{\label{fig:cellB}\includegraphics[width=0.17\textwidth]{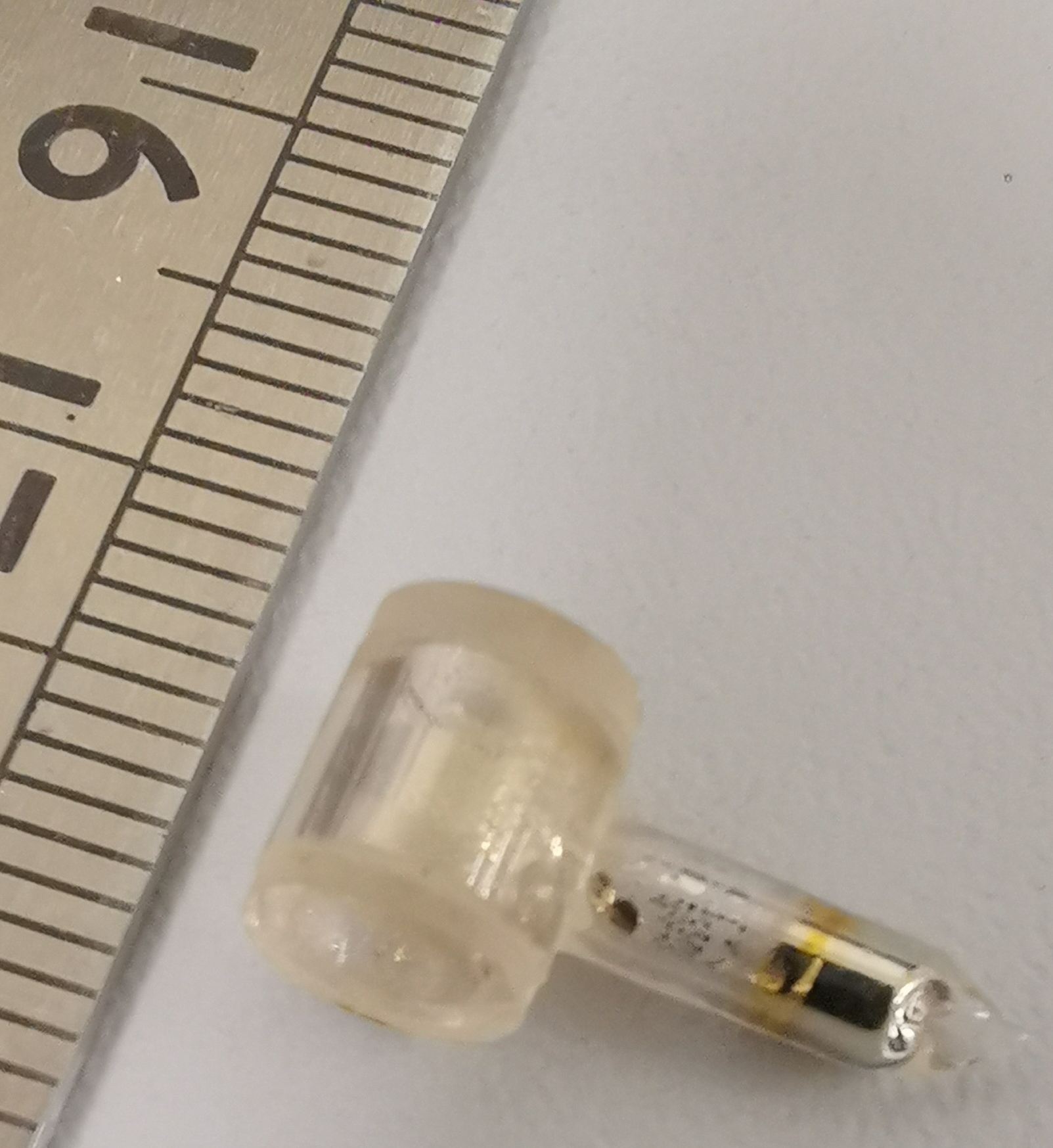}}
\subfigure[]{\label{fig:cellC}\includegraphics[width=0.138\textwidth]{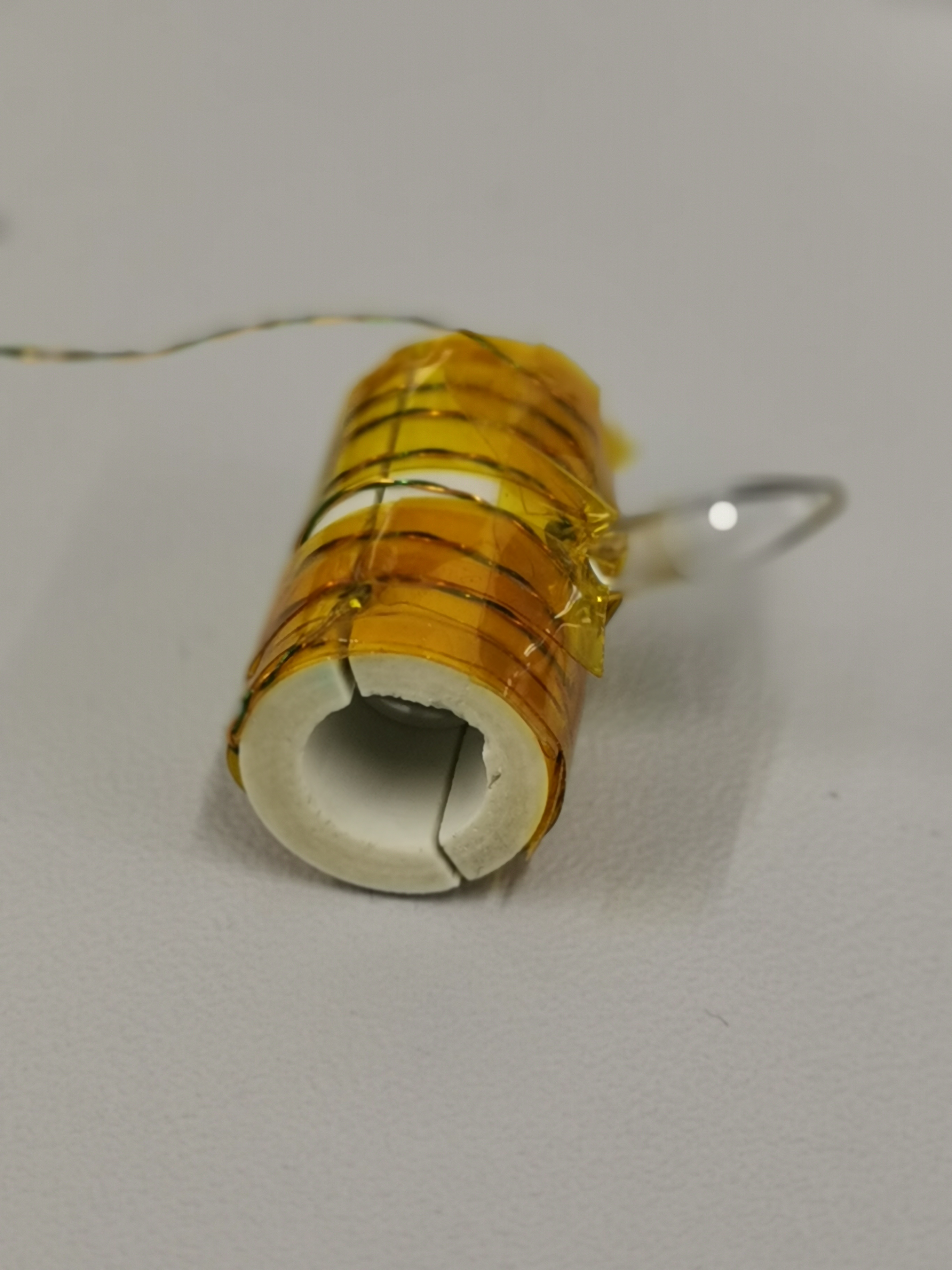}}
\caption{\label{fig:}(a) Schematic of an alignment-based magnetometer. The laser light propagates along the $y$-direction and is $z$-polarised. Components include: half-wave plates ($\lambda/2$), polarising beam splitters (PBS), a vapour cell (Cell) and a balanced photodetector (BPD). Static $\textbf{B}_{0}=B_{0}\hat{\textbf{z}}$ and oscillating magnetic fields $\textbf{B}_{\text{RF}}(t)=B_{\text{RF}}(t)\hat{\textbf{x}}$ are applied at the position of the vapour cell. (b) Photo of the paraffin-coated cell. (c) Photo of the buffer gas cell. (d) Photo of the buffer gas cell surrounded by a Shapal ceramic cylinder, heating wires and Kapton tape.}
\end{figure}

A schematic of the experimental setup is shown in Fig.~\ref{fig:alignment_magnetometer_setup}. A diode laser system outputs light resonant with the $F=4\rightarrow F'=3$  Cs D1 transition (895~nm). The light is passed through an optical fiber and is collimated at its output. The linearly polarised light with an electric field amplitude $E_{0}\hat{\textbf{z}}$ then passes through 
a cubic (5~mm)$^{3}$ hand-blown paraffin-coated vapour cell (see Fig.~\ref{fig:cellA}). The vapour cell is kept at room temperature ($\sim18.5\degree$C) and placed inside a magnetic shield (Twinleaf MS-1). Static $B_0\hat{\textbf{z}}$ and oscillating 
$B_{\text{RF}}\cos(2\pi \nu_{\text{RF}}t)\hat{\textbf{x}}$ 
magnetic fields can be applied using coils inside the magnetic shield.  Here $\nu_{\text{RF}}$ is the RF frequency in Hz, while $\omega_{\text{RF}}=2\pi\nu_{\text{RF}}$  is the RF frequency in rad/s.
Polarimetry is then performed using a half-wave plate, a polarising beam splitter, and a balanced photodetector (Thorlabs PDB210A/M) to detect the polarization rotation of the transmitted light. The resultant photodetector voltage is demodulated at the RF frequency $\nu_{\text{RF}}$ using a lock-in amplifier (SR830) such that in-phase $X$ and out-of-phase $Y$ signals are obtained.
\iffalse
\begin{figure}
\includegraphics[width=\linewidth]{AlignmentMagnetometer (1).pdf}
\caption{Schematic of an alignment-based magnetometer. The laser light propagates along the $y$-direction and is $z$-polarised. Components include half-wave plates ($\lambda/2$), polarising beam splitters (PBS), a paraffin-coated or buffer gas Cs vapour cell (Cell), a balanced photodetector (BPD), and static $\textbf{B}_{0}=B_{0}\hat{\textbf{z}}$ and oscillating magnetic fields $\textbf{B}_{\text{RF}}(t)=B_{\text{RF}}(t)\hat{\textbf{x}}$ at the position of the vapour cell.  \label{fig:}
}
\end{figure}
\fi

\begin{figure}[ht]
\includegraphics[width=\linewidth]{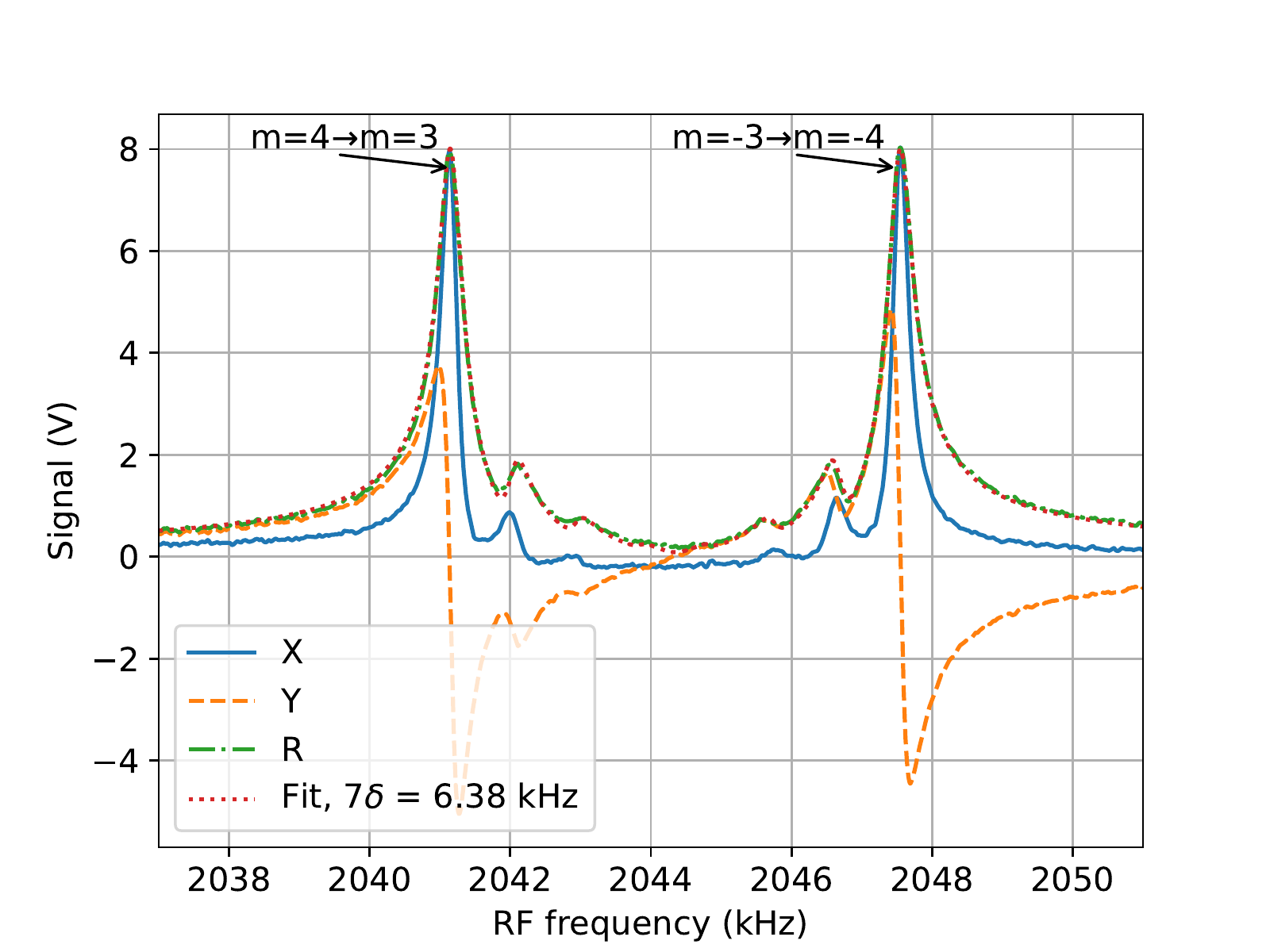}
\caption{
Non-linear Zeeman splitting of the magnetic resonances using a paraffin-coated cell. The magnitude $R$ is fitted to Eq.~\ref{eq:fit}. The fit is included as a dotted line. The magnetic resonances for $m=4\rightarrow m=3$ and $m=-3\rightarrow m=-4$, with different Larmor frequencies, are indicated.
\label{fig:AM_Paraffin_NonLinearZeemanSplitting}
}
\end{figure}

The optical pumping of an aligned state can be experimentally verified by exploiting the non-linear Zeeman effect. 
These measurements were done at a relatively large static magnetic field ($B_0=5.84$~G) corresponding to a Larmor frequency close to 2~MHz.
When the RF frequency was swept over the range 2.037-2.051~MHz, we observe a magnetic resonance spectrum with several peaks  (see  Fig.~\ref{fig:AM_Paraffin_NonLinearZeemanSplitting}). 
The two largest peaks correspond to the transitions
$m=4\rightarrow m=3$ and $m=-3\rightarrow m=-4$ with
transition frequencies
$\Delta \nu_{4,3}$ and $\Delta \nu_{-3, -4}$,  respectively.
The difference in transition frequencies 
$|\Delta \nu_{4,3} -\Delta \nu_{-3, -4}|$ is experimentally found to be $6.38(0.02)$~kHz, agreeing with the value $7\delta = 6.37$~kHz
calculated from Eqs.~\ref{eq:NZS0} and \ref{eq:NZS}, confirming that we are observing the non-linear Zeeman splitting. This difference in transition frequencies was extracted by fitting the data of $R$ in Fig.~\ref{fig:AM_Paraffin_NonLinearZeemanSplitting} to the function \cite{julsgaard_2003}

\begin{equation}
    \text{R}=\left|\sum_{m=-3}^{4} 
    \frac{ A_{m, m-1} \left[ 1 + i\left(\nu_{\text{RF}}-\nu_{m, m-1} \right)/\Tilde{\gamma} \right] }{\left( \nu_{\text{RF}}-\nu_{m, m-1} \right)^2 + \Tilde{\gamma}^2 }
    \right|.
    \label{eq:fit}
\end{equation}
which is a sum of eight magnetic resonances  with resonance frequencies 
$\nu_{m,m-1}=\nu_L-\delta \left( m -\frac{1}{2}\right)$ and half width at half maximum (HWHM) $\Tilde{\gamma}=1/(2\pi T_{2})$ (in Hz) as seen by comparison with Eq.~\ref{eq:R} and illustrated in Fig.~\ref{fig:Threelevels}. 
The data was fitted with seven free parameters: four amplitudes $A_{4,3}$, $A_{3,2}$, $A_{2,1}$, $A_{1,0}$ (as the magnetic resonance spectrum is symmetric such that $A_{0,-1}=A_{1,0}$, $A_{-1,-2}=A_{2,1}$, $A_{-2,-3}=A_{3,2}$, $A_{-3,-4}=A_{4,3}$), the Larmor frequency $\nu_L$, the non-linear Zeeman splitting $\delta$, and the width $\Tilde{\gamma}$.

In total, the spectrum has eight peaks, although the middle two are hardly visible in Fig.~\ref{fig:AM_Paraffin_NonLinearZeemanSplitting} due to their smaller height.
The height of the individual peaks corresponding to $A_{m,m-1}/\Tilde{\gamma}^2$ in Eq.~\ref{eq:fit}  are proportional to the difference in populations of neighbouring magnetic sublevels \cite{julsgaard_2003}. This is why there are eight peaks in the non-linear Zeeman splitting, but nine populations in Fig.~\ref{fig:chapter02_Pi_OPvstime_BarGraph_AM}. 
As the outermost peaks are largest and have equal height, we conclude that an aligned state is created in the $F=4$ ground state, with the majority of the atoms pumped into the $F=4, m=\pm4$ states. 
The optical pumping is not perfect as some of the atoms are pumped into the other magnetic sublevels. This is due to the non-zero longitudinal relaxation rate $\Gamma_{1}$.
\iffalse
 \begin{figure*}
\subfigure[]{\label{fig:AM_Paraffin_Sensitivity_10kHz_Thorlabs_RFSweep}\includegraphics[width=0.49\textwidth]{AM_Paraffin_Sensitivity_10kHz_Thorlabs_RFSweep (1).pdf}}
\subfigure[]{\label{fig:AM_Paraffin_Sensitivity_10kHz_Thorlabs_RFOFF}\includegraphics[width=0.49\textwidth]{AM_Paraffin_Sensitivity_10kHz_Thorlabs_RFOFF (2).pdf}}
\caption{\label{fig:}Sensitivity measurement of a paraffin-coated alignment-based magnetometer ($T\sim20\degree$C) at a Larmor frequency of $\nu_L=10.25$~kHz. (a) Magnetic resonance with the RF frequency swept between 9 and 11.5~kHz. (c) A 240~s time trace with the RF field turned off and with the lock-in amplifier demodulating signals at $\nu_L$.}
\end{figure*}
\fi

\begin{figure*}
\subfigure[]{\label{fig:AM_Paraffin_Sensitivity_10kHz_Thorlabs_RFSweep}\includegraphics[width=0.49\textwidth]{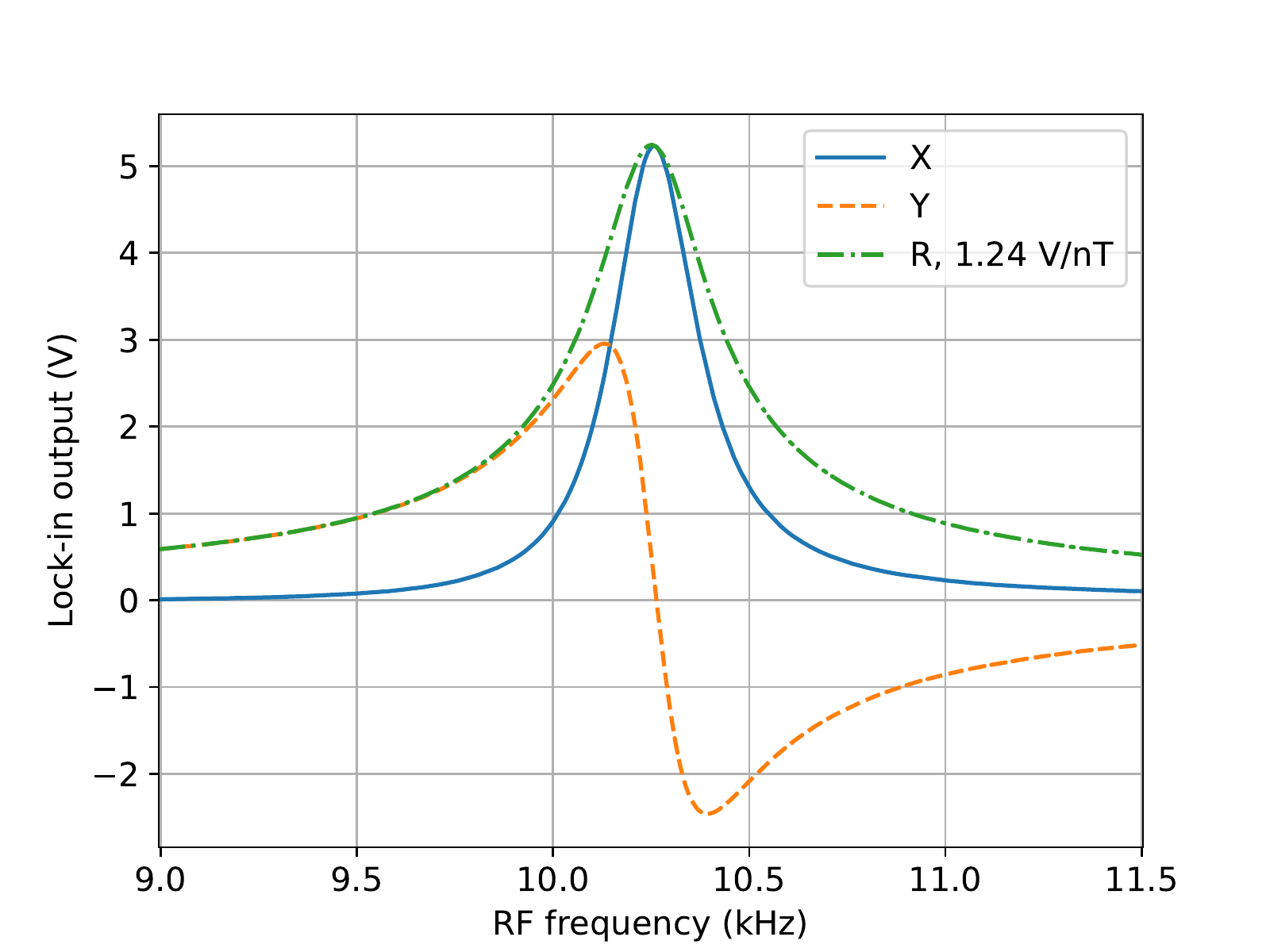}}
\subfigure[]{\label{fig:AM_Paraffin_Sensitivity_10kHz_Thorlabs_RFOFF}\includegraphics[width=0.49\textwidth]{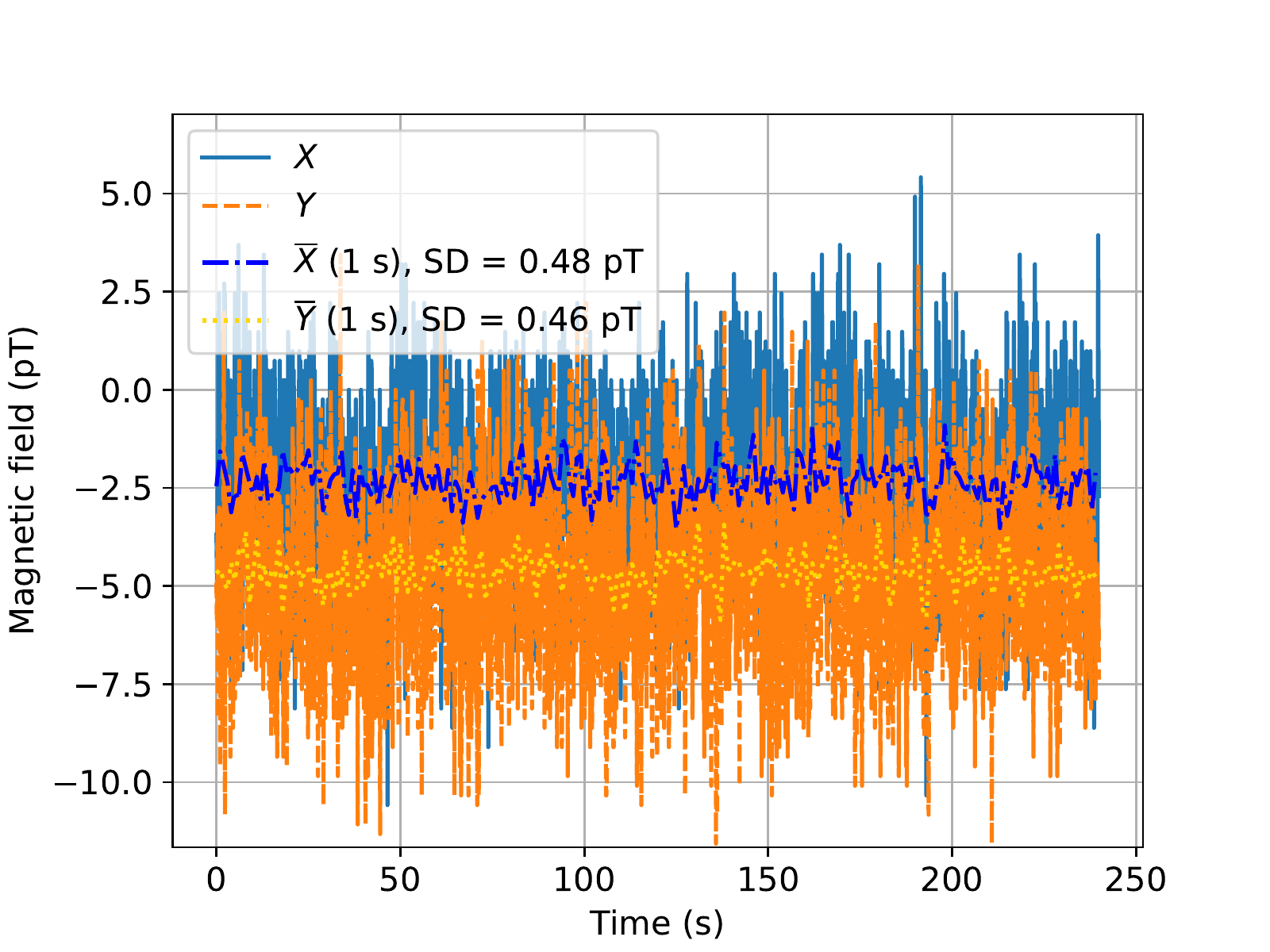}}
\subfigure[]{\label{fig:AM_100Torr_RFSweep}\includegraphics[width=0.49\textwidth]{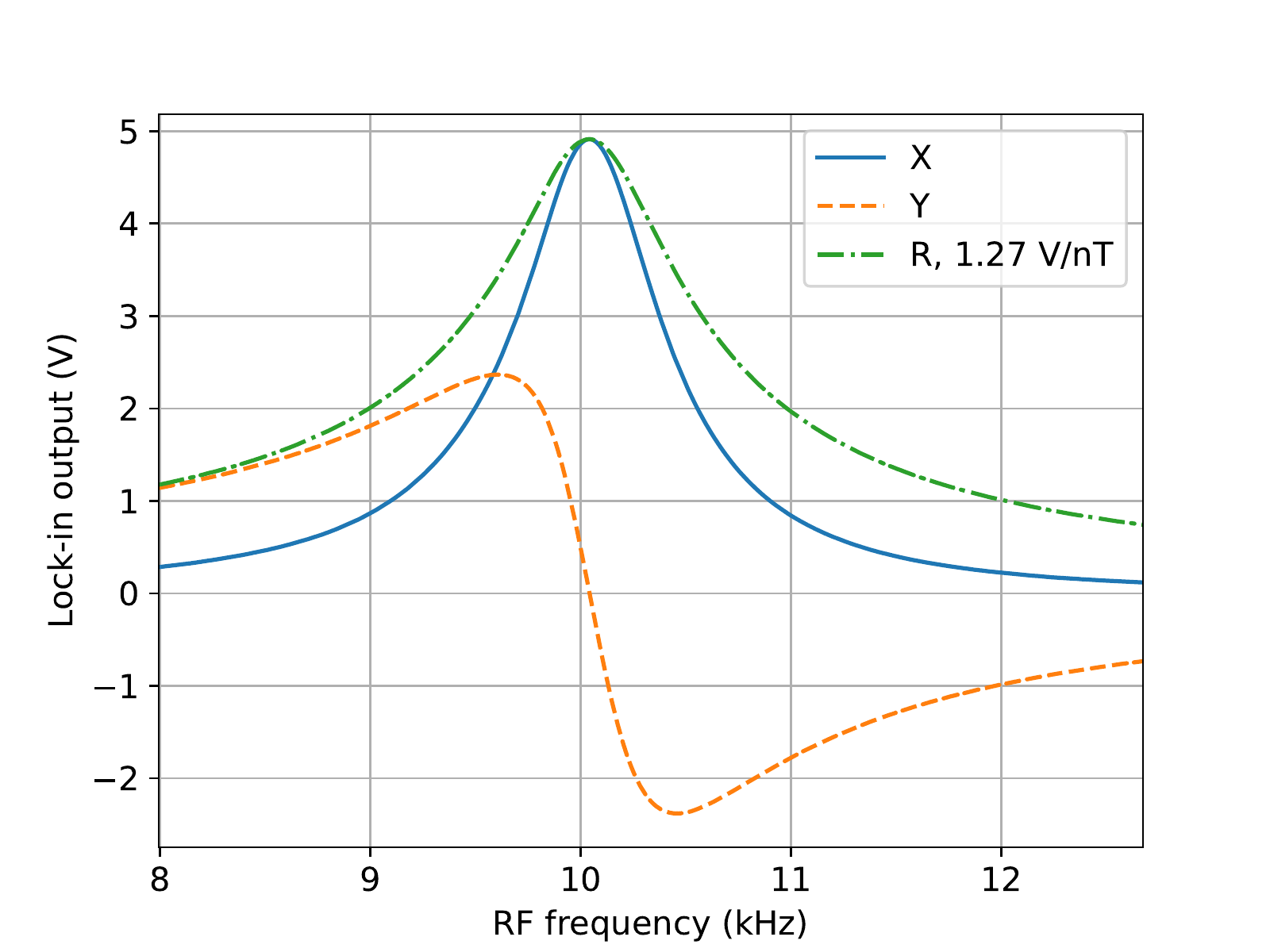}}
\subfigure[]{\label{fig:AM_100Torr_RFOFF}\includegraphics[width=0.49\textwidth]{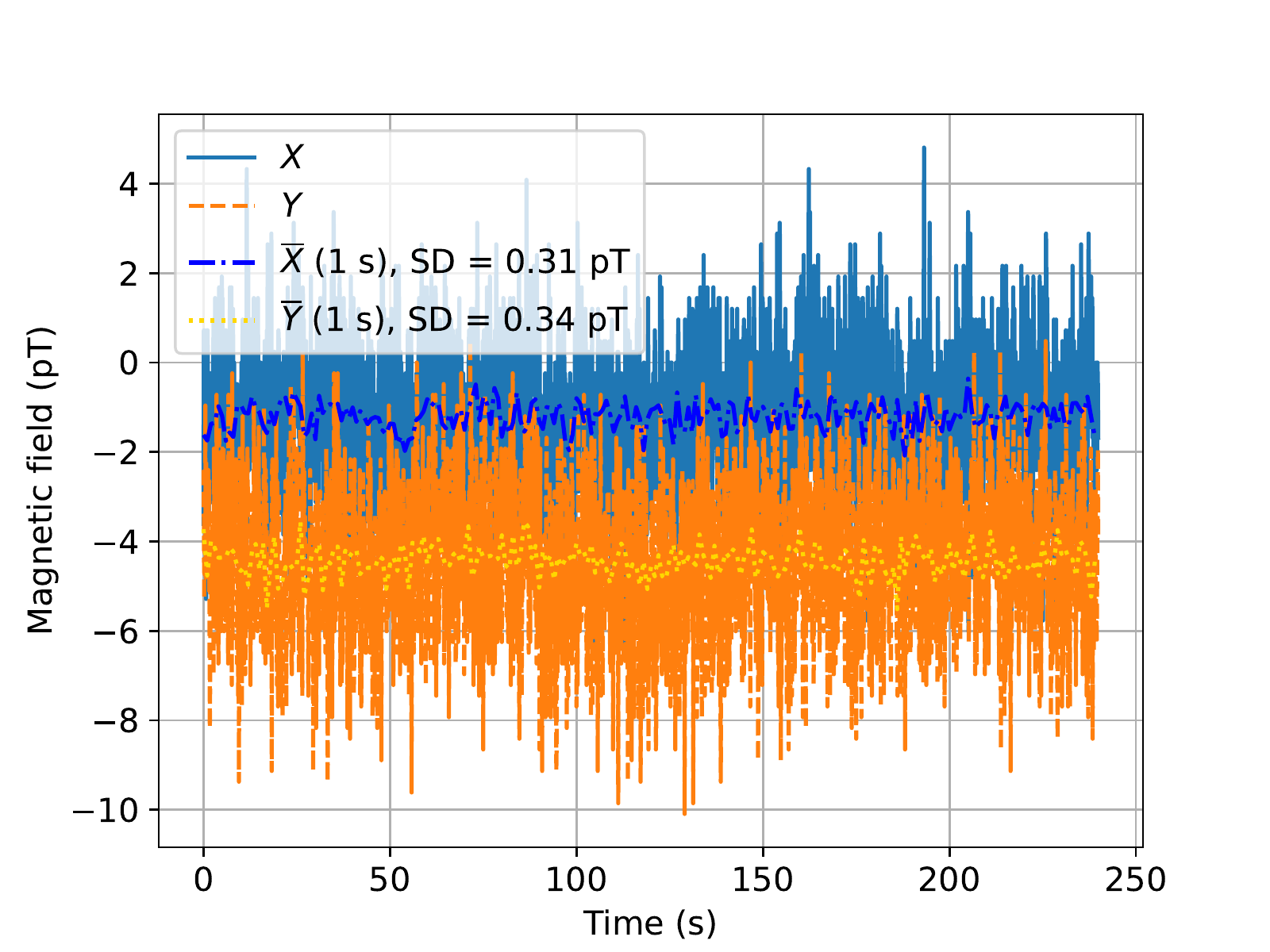}}
\caption{\label{fig:}(a) (b) Sensitivity measurement of the paraffin-coated alignment-based magnetometer ($T\sim20\degree$C) at a Larmor frequency of $\nu_L=10.25$~kHz. (a) Magnetic resonance with the RF frequency swept between 9 and 11.5~kHz. (b) A 240~s time trace of the intrinsic OPM noise with the lock-in amplifier demodulating signals at $\nu_L$. (c) (d) Sensitivity measurement of the 65~Torr N$_{\text{2}}$ ($T\sim55\degree$C) at a Larmor frequency of $\nu_{L}=10.04$~kHz. (c) Magnetic resonance with the RF frequency swept between 8 and 12.5~kHz. (d) A 240~s time trace of the intrinsic OPM noise with the lock-in amplifier demodulating signals at $\nu_L$.}
\end{figure*}

We now proceed with characterising the magnetic field sensitivity of the paraffin-coated vapour cell. These measurements were carried out at a smaller static magnetic field $B_0$ corresponding to a Larmor frequency of around 10~kHz.
A $10~\mu$W light power beam passed through the cell and a 4.22~nT$_{\text{RMS}}$ (20~mV$_{\text{RMS}}$) oscillating magnetic field was applied. A magnetic resonance signal is shown in  Fig.~\ref{fig:AM_Paraffin_Sensitivity_10kHz_Thorlabs_RFSweep}, where the RF frequency was swept between 9~kHz and 11.5~kHz and $\nu_{L}=10.25$~kHz. From this, the peak of the resonance signal is extracted and divided by the applied oscillating magnetic field to give a conversion between the lock-in amplifier readout and the corresponding RF field amplitude $B_{\text{RF}}$. Once this calibration was completed in Fig.~\ref{fig:AM_Paraffin_Sensitivity_10kHz_Thorlabs_RFSweep}, the lock-in demodulation frequency was fixed to the Larmor frequency, the RF amplitude was set to zero ($B_{\text{RF}}=0$), and a 4~minute time trace of the intrinsic noise of the OPM taken (see  Fig.~\ref{fig:AM_Paraffin_Sensitivity_10kHz_Thorlabs_RFOFF}). Following on from this the light hitting the balanced photodetector was completely blocked and another time trace obtained (data not shown). The sensitivity to small oscillating magnetic fields, i.e., the intrinsic OPM noise, is 480~fT/$\sqrt{\text{Hz}}$ for $X$ and 460~fT/$\sqrt{\text{Hz}}$ for $Y$. This was calculated using the methods described in Ref.~\cite{rushton_2022} by taking the standard deviations (SDs) of 240$\times$1~s averaged segments, which are included in the legend of Fig.~\ref{fig:AM_Paraffin_Sensitivity_10kHz_Thorlabs_RFOFF}.
The SDs with the light blocked are only just below at 410~fT/$\sqrt{\text{Hz}}$ for $X$ and 380~fT/$\sqrt{\text{Hz}}$ for $Y$. 
This noise is mainly due to electronic noise of the balanced photodetector and also due to a small contribution from the electronic noise of the data-acquisition system. 
The signal size and thereby the sensitivity could be improved by heating the vapour cell \cite{ledbetter_acosta_rochester_budker_pustelny_yashchuk_2007, seltzer_thesis} and using a larger vapour cell.
A fundamental limit to the sensitivity is given by the spin-projection noise
\cite{ledbetter_acosta_rochester_budker_pustelny_yashchuk_2007, graf_kimball_rochester_kerner_wong_budker_alexandrov_balabas_yashchuk_2005}
\begin{equation}
\delta B_{\text{spn}}=\frac{2\hbar}{g_{F} \mu_{B}\sqrt{n V T_{2}}},
\label{eq:spinprojectionnoise}
\end{equation}  
where $g_{F}=1/4$ for the $F=4$ Cs ground state, $n\sim 2.2\times10^{16}$~m$^{-3}$ ($T\sim18.5\degree$C) is the number density of Cs atoms, $T_{2}\sim1/(\pi (230~\text{Hz}))\sim 1.4$~ms is the transverse relaxation time and $V=(5~\text{mm})^{3}$ is the volume of the whole cell, as all the atoms in the cell are probed. The sensitivity is estimated to be $\delta B_{\text{spn}}\sim50$~fT/$\sqrt{\text{Hz}}$ using the numbers above. A balanced photodetector with reduced electronic noise would help us get closer to this quantum-limited sensitivity.

\section{Buffer gas cell}
\label{sec:buffer}
We now carry out experiments with a hand-blown cylindrical buffer gas cell (5~mm length, 5~mm diameter) filled with Cs as well as N$_{\text{2}}$ buffer gas  (see Fig.~\ref{fig:cellB}).
The buffer gas cell is surrounded by a Shapal ceramic cylinder, which is chosen for its high thermal conductivity. The ceramic cylinder is wrapped in a non-magnetic resistive twisted wire and wrapped with heat insulator aerogel and Kapton tape as shown in  Fig.~\ref{fig:cellC}. 
The buffer gas cell can then be heated and kept at an elevated temperature by running current through the twisted wire.

\begin{figure}
\includegraphics[width=\linewidth]{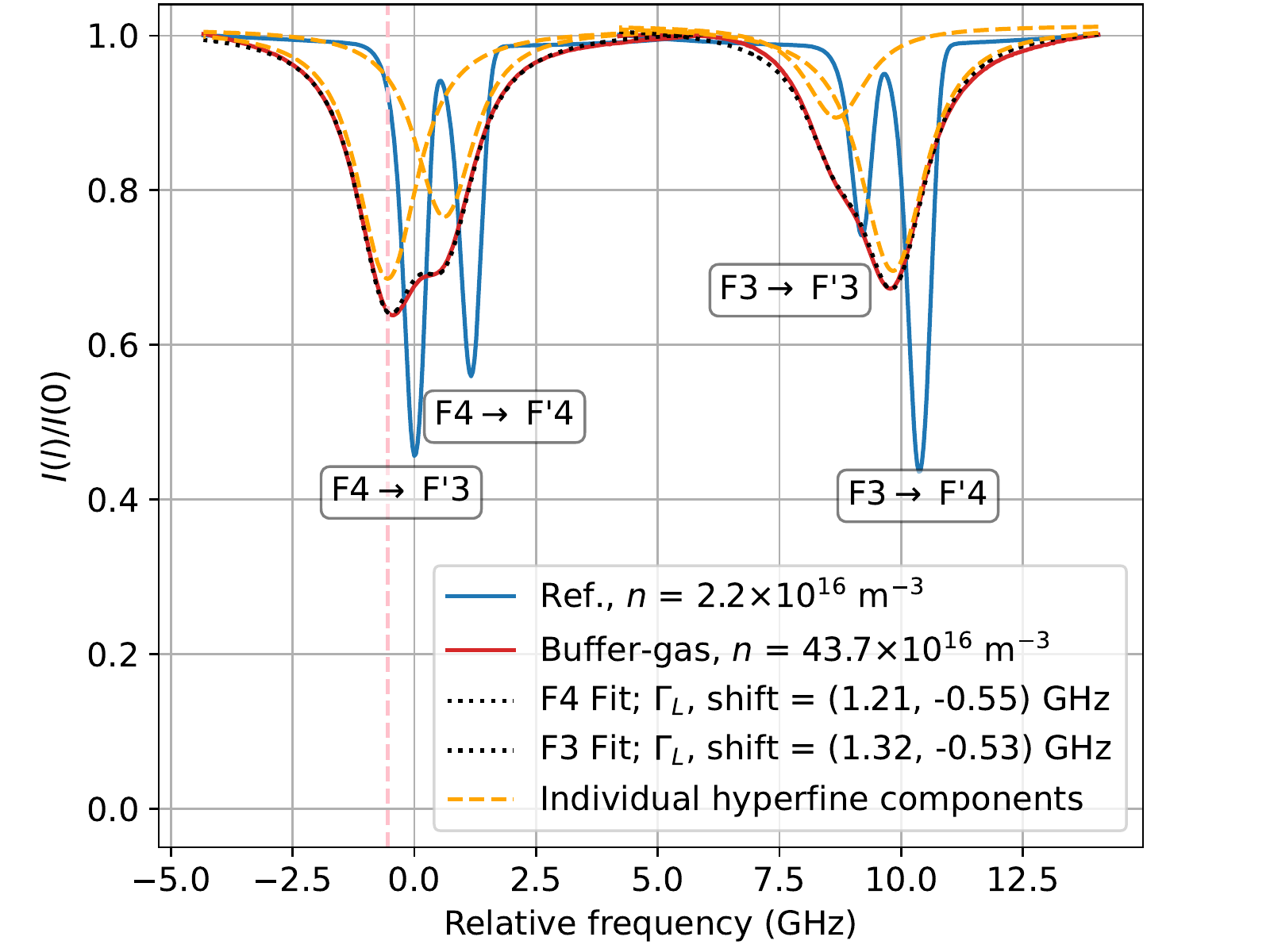}
\caption{Absorption spectrum of the D1 line with a 65(3)~Torr N$_{\text{2}}$ cell alongside a frequency reference which is a pure Cs cell. The buffer gas cell is heated to 51$\degree$C corresponding to a density of 43.7$\times10^{16}$~m$^{-3}$ Cs atoms and a Doppler linewidth $\Gamma_{G}=374$~MHz.
%by applying 300~mA to the heater. 
The $F=3\rightarrow F'=3,4$ and $F=4\rightarrow F'=3,4$ transitions are fitted to Voigt profiles and the Lorentzian width $\Gamma_{L}$ and pressure shift are extracted. \label{fig:ABM_100TorrAbsorptionSpectroscopy_intro}
}
\end{figure}

The N$_{\text{2}}$ buffer gas pressure was determined using  absorption spectroscopy as described by Andalkar \cite{andalkar_warrington_2002}. 
The laser power was kept low for these absorption measurements to avoid any optical pumping effects.
An absorption spectrum of the buffer gas cell is obtained, plotted on top of an absorption spectrum of a pure Cs cell (75~mm length and kept at room-temperature) in Fig.~\ref{fig:ABM_100TorrAbsorptionSpectroscopy_intro}. The pure cell only contains Cs (and neither contains paraffin or buffer gas) and is used as a frequency reference.
The absorption spectrum for the pure cell shows four absorption resonances separated by ground and excited state hyperfine splittings (9.2~GHz, 1.2~GHz) as expected for Cs D1 spectroscopy.
The absorption resonances have a Voigt lineshape, which is a convolution of a Lorentzian and Gaussian lineshape. 
For the pure cell, the Gaussian Doppler width is much larger than the Lorentzian natural linewidth 4.6~MHz full width at half maximum (FWHM) of the Cs excited state.
For a buffer gas cell, collisions between buffer gas atoms and Cs atoms lead to Lorentzian pressure broadening as well as frequency shifts of the absorption resonances, as seen in Fig.~\ref{fig:ABM_100TorrAbsorptionSpectroscopy_intro}.
The pressure broadening is extracted by fitting the $F=3\rightarrow F'=3$ and $F=3\rightarrow F'=4$ absorption resonances to a sum of two Voigt profiles and using their relative hyperfine strengths (1/4 and 3/4, respectively) and then repeating the procedure for $F=4\rightarrow F'=3$ and $F=4\rightarrow F'=4$, with hyperfine strengths of 7/12 and 5/12, respectively. 
The Doppler width $\Gamma_{G}$ is fixed (374~MHz FWHM at 51$\degree$C) and the Lorentzian $\Gamma_{L}$ (1.26(0.05)~GHz) is fitted, corresponding to a pressure of 65(3)~Torr, using the conversion of 19.51~MHz/Torr from \cite{andalkar_warrington_2002} for the D1 pressure broadening with N$_{\text{2}}$. The pressure can also be extracted from the shift -~0.54(0.01)~GHz in resonance frequencies, which corresponds to a pressure of 65(1)~Torr. 

Our alignment-based magnetometer uses $\pi$-polarized light resonant with the $F=4\rightarrow F'=3$ transition (see Fig.~\ref{fig:Threelevels}(b)), as in this case, the $F=4,m=\pm4$ states are dark states and atoms become optically pumped into those states with equal probability, creating the spin-alignment, as depicted in Fig.~\ref{fig:chapter02_Pi_OPvstime_BarGraph_AM}b. Note that for $\pi$-polarized light resonant with the $F=4\rightarrow F'=4$ transition, the $F=4,m=0$ sublevel will be a dark state instead.
With buffer gas pressure broadening, 
the  $F=4\rightarrow F'=3$ and $F=4\rightarrow F'=4$ resonances begin to overlap. From our fit, we deduce that the overlap is only $\sim 10\%$ for our pressure of 65 Torr N$_{\text{2}}$ (see Fig.~\ref{fig:ABM_100TorrAbsorptionSpectroscopy_intro} and the thin dotted vertical line). 
At higher pressures the two transitions will overlap even more. This is problematic for an alignment-based magnetometer as the light in this case will drive both $F=4\rightarrow F'=3$ and $F=4\rightarrow F'=4$ transitions at the same time. The $F=4,m=\pm=4$ are then not dark states and significantly less spin-alignment is created. 

\begin{figure}
\includegraphics[width=\linewidth]{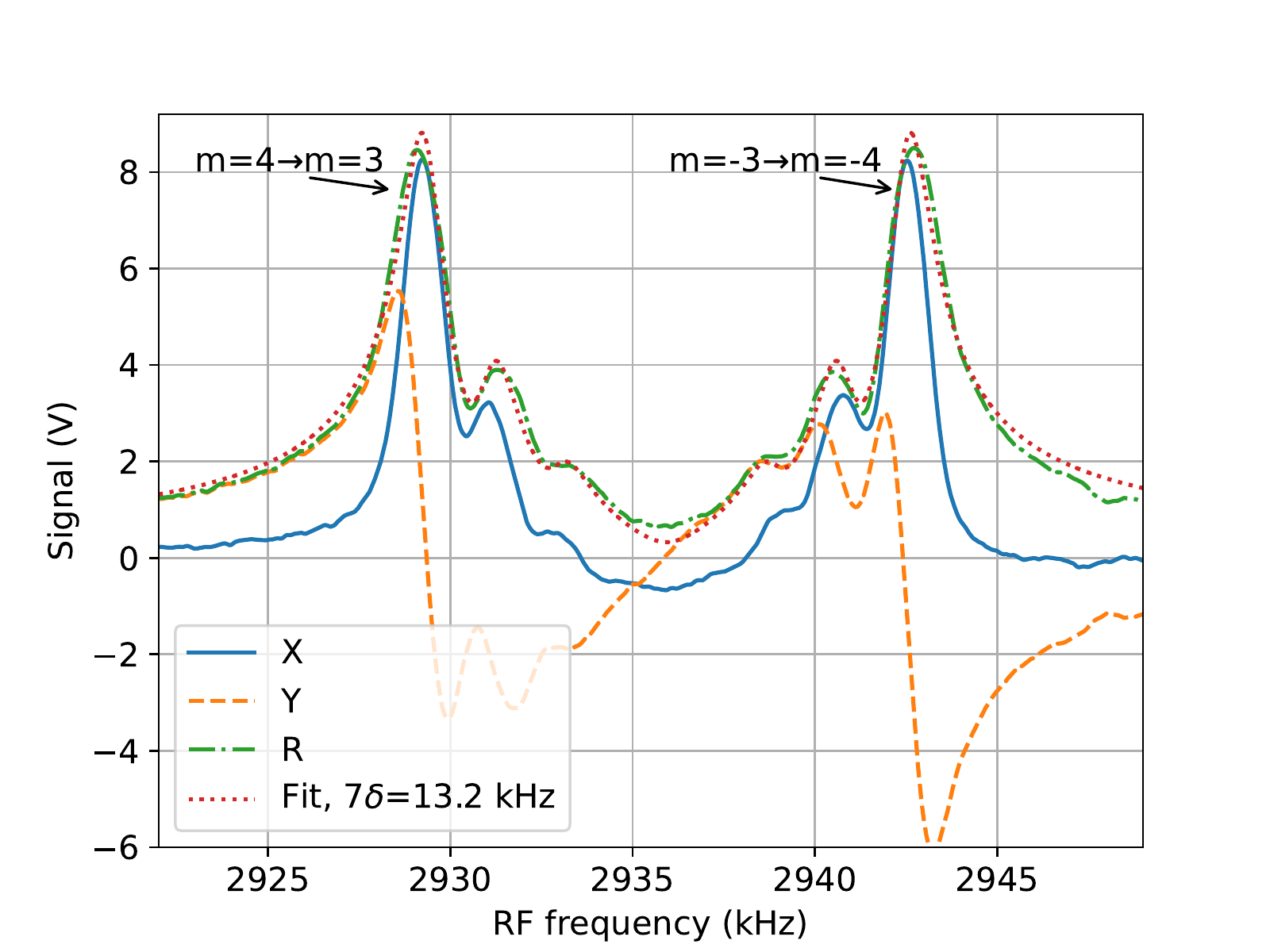}
\caption{
Non-linear Zeeman splitting of the magnetic resonances using a 65~Torr N$_{\text{2}}$ buffer gas cell heated to $\sim 55\degree$C. The magnitude $R$ is fitted to Eq.~\ref{eq:fit}. The magnetic resonances for $m=4\rightarrow m=3$ and $m=-3\rightarrow m=-4$ are indicated.
}\label{fig:NZS_100Torr}
\end{figure}

To verify whether optical pumping into the $F=4, m=\pm4$ states is possible with the 65~Torr N$_{\text{2}}$ buffer gas cell where the excited hyperfine states partially overlap ($\sim 10\%$) and where quenching is the main de-excitation mechanism as described previously,  
once again the static field is adjusted to be large ($B_0=8.38$~G) and a magnetic resonance spectrum is recorded (see Fig.~\ref{fig:NZS_100Torr}).
Again we see the magnetic resonances split due to the non-linear Zeeman effect, and the two outermost resonances have the largest and equal heights.
The frequency difference between the $m=4\rightarrow m=3$ transition and the $m=-3\rightarrow m=-4$ transition 
is found experimentally to be $|\Delta \nu_{4,3}-\Delta \nu_{-3,-4}|=13.2(0.1)$~kHz from a fit of the data in Fig.~\ref{fig:NZS_100Torr} to Eq.~\ref{eq:fit}, which agrees well with the value $7\delta=13.1$~kHz calculated from Eqs.~\ref{eq:NZS0} and \ref{eq:NZS}. 

This experimentally demonstrates  that it is possible to generate a spin aligned stated in the 65~Torr N$_{\text{2}}$ buffer gas cell by optically pumping more Cs atoms into the $m=\pm4$ states than the other magnetic sublevels in the $F=4$ ground state. 
It is expected that better optical pumping into the $m=\pm4$ states will be achieved if a smaller buffer gas pressure is used, as there will be less unwanted pumping to the $F=4\rightarrow F'=4$ transition. A higher ratio $R_p/\Gamma_{1}$ (see Eq.~\ref{eq:rateBuffer}) will also increase pumping into the $m=\pm4$ states. The drawback of a lower buffer gas pressure, however, is that the atoms will diffuse more quickly to the walls, leading to a smaller $T_{2}$ time and hence a less sensitive OPM. These two processes compete and need to be taken into consideration when selecting the optimal buffer gas pressure for an alignment-based magnetometer.

\iffalse
 \begin{figure*}
\subfigure[]{\label{fig:AM_Paraffin_Sensitivity_10kHz_Thorlabs_RFSweep}\includegraphics[width=0.49\textwidth]{AM_Paraffin_Sensitivity_10kHz_Thorlabs_RFSweep (1).pdf}}
\subfigure[]{\label{fig:AM_Paraffin_Sensitivity_10kHz_Thorlabs_RFOFF}\includegraphics[width=0.49\textwidth]{AM_Paraffin_Sensitivity_10kHz_Thorlabs_RFOFF (2).pdf}}
\caption{\label{fig:}Sensitivity measurement of a paraffin-coated alignment-based magnetometer ($T\sim20\degree$C) at a Larmor frequency of $\nu_L=10.25$~kHz. (a) Magnetic resonance with the RF frequency swept between 9 and 11.5~kHz. (c) A 240~s time trace with the RF field turned off and with the lock-in amplifier demodulating signals at $\nu_L$.}
\end{figure*}
\fi

We now characterise the magnetic field sensitivity of the buffer gas cell using the same procedure which was used for the paraffin-coated cell.
The optimal light power was found to be 30~$\mu$W. 
A magnetic resonance signal at 10~kHz was obtained with the 65 N$_{\text{2}}$  Torr cell in Fig.~\ref{fig:AM_100Torr_RFSweep}. A 240~s time trace with the RF field turned off is shown in Fig.~\ref{fig:AM_100Torr_RFOFF}. The sensitivity of the OPM, defined as the SD of the 240$\times 1$~s data points in Fig.~\ref{fig:AM_100Torr_RFOFF}, is 310~fT/$\sqrt{\text{Hz}}$ for $X$  and 340~fT/$\sqrt{\text{Hz}}$ for $Y$. The sensitivity of the buffer gas cell therefore exceeds the paraffin-coated cell in this paper.
The sensitivity is mainly limited by laser shot noise and electronic noise of the balanced photodetector.

We use Eq.~\ref{eq:spinprojectionnoise} to calculate the predicted quantum-limited spin-projection noise. The number density $n=60\times10^{16}$~m$^{-3}$ at $T=55\degree$C and $T_{2}=1/(\pi (800~\text{Hz}))$. In a buffer gas cell only the atoms inside the beam are probed, unlike in a paraffin-coated cell where all the atoms in the cell are probed. 
We therefore use the volume inside the beam $V=V_{\text{beam}}=3.9\times10^{-9}$~m$^{3}$, where the diameter of the beam is $\sim 1$~mm and length of the cell is 5~mm.
Inserting the numbers above, we estimate the atomic noise to be $\delta B_{\text{spn}}\sim~$100~fT/$\sqrt{\text{Hz}}$.
A better sensitivity could be obtained by increasing the diameter and length of the cell, whilst increasing the size of the beam. If a 5~mm diameter beam was used, probing the whole cell, the atomic noise is estimated to be $\delta B_{\text{spn}}\sim20$~fT/$\sqrt{\text{Hz}}$. Note that many atoms are lost to the $F=3$ ground state (see Fig.~\ref{fig:chapter02_Pi_OPvstime_BarGraph_AM}), reducing the number of Cs atoms that are probed. Using a second laser beam (typically called a repumper) bringing the atoms out of $F=3$ and back into $F=4$ would also increase the number of probed atoms, improving the sensitivity of the RF OPM.

\section{Conclusions} 
The results presented in this paper demonstrate the first implementation of a one-beam radio-frequency optically pumped magnetometer (RF OPM), the alignment-based magnetometer, being used with a buffer gas cell. 
The sensitivity of the alignment-based magnetometer with Cs alkali vapour and 65~Torr N$_{\text{2}}$ buffer gas  was $325$~fT/$\sqrt{\text{Hz}}$. 
This sensitivity could be further improved upon by using a balanced photodetector with lower electronic noise.
Further studies could investigate the optimal vapour cell size, operating temperature and buffer gas pressure.
Although our experiments were carried out using hand-blown vapour cells, we expect similar performance with microfabricated buffer gas cells. 
Our work opens up the possibility of the commercialisation of compact, robust and portable RF OPMs using only one laser beam with buffer gas cells, a much more scalable and commercially viable option than using paraffin-coated vapour cells. 
%

%TC:ignore
\begin{acknowledgments}
This work was supported by the UK Quantum Technology Hub in Sensing and Timing, funded by the Engineering and Physical Sciences Research Council (EPSRC) (Grant No. EP/T001046/1), the QuantERA grant C’MON-QSENS! by EPSRC (Grant No. EP/T027126/1), the Nottingham Impact Accelerator/EPSRC Impact Acceleration Account (IAA),  and the Novo Nordisk Foundation (Grant No. NNF20OC0064182). We thank Janek Kolodynski and Marcin Koźbiał for reading and commenting on the manuscript.
\end{acknowledgments}

\section*{Data Availability Statement}
Further data can be available from the authors upon request.
%TC:endignore
%\newpage

%TC:ignore
\appendix

\end{document}